\documentclass[10pt,journal,twoside]{IEEEtran}
\usepackage{amsmath,amssymb,mathtools,url}
\usepackage{graphicx,cite}
\usepackage{epstopdf}
\usepackage{hyperref}

\newtheorem{definition}{\textbf{Definition}}
\newtheorem{problem}{\textbf{Problem}}
\newtheorem{theorem}{\textbf{Theorem}}
\newtheorem{remark}{\textbf{Remark}}
\newtheorem{lemma}{\textbf{Lemma}}

\begin{document}

\title{Spectral and Energy Efficiency Trade-Offs in Cellular Networks}

\author{Dimitrios~Tsilimantos,~\IEEEmembership{Member,~IEEE,}
        Jean-Marie~Gorce,~\IEEEmembership{Senior Member,~IEEE,}
        Katia~Jaffr\`es-Runser,~\IEEEmembership{Member,~IEEE}
        and~H.~Vincent~Poor,~\IEEEmembership{Fellow,~IEEE,} 
\thanks{This work was partly produced in the framework of the common research lab between INRIA and Alcatel-Lucent Bell labs and presented in the framework of the GreenTouch initiative \cite{GreenTouch}.}
\thanks{D. Tsilimantos was with INRIA, University of Lyon, CITI-INRIA, Villeurbanne F-69621, France. He is now with Huawei Technologies, France Research Center, Mathematical and Algorithmic Sciences Lab, 20 Quai du Point du Jour, 92100, Boulogne-Billancourt (e-mail: \mbox{dimitrios.tsilimantos@huawei.com}).}
\thanks{J-M. Gorce was with Princeton University, Princeton, NJ, USA. He is now with INRIA, University of Lyon, CITI-INRIA, Villeurbanne F-69621, France (e-mail: jean-marie.gorce@insa-lyon.fr).}
\thanks{K. Jaffr\`es-Runser is with University of Toulouse, IRIT, INPT-ENSEEIHT, Toulouse 31071, France (e-mail: katia.jaffres-runser@irit.fr).}
\thanks{H. V. Poor is with the Department of Electrical Engineering, Princeton University, Princeton, NJ, 08544, USA (e-mail: poor@princeton.edu).}
}

\maketitle
\begin{abstract}
This paper presents a simple and effective method to study the spectral and energy efficiency (SE-EE) trade-off in cellular networks, an issue that has attracted significant recent interest in the wireless community. The proposed theoretical framework is based on an optimal radio resource allocation of transmit power and bandwidth for the downlink direction, applicable for an orthogonal cellular network. The analysis is initially focused on a single cell scenario, for which in addition to the solution of the main SE-EE optimization problem, it is proved that a traffic repartition scheme can also be adopted as a way to simplify this approach. By exploiting this interesting result along with properties of stochastic geometry, this work is extended to a more challenging multi-cell environment, where interference is shown to play an essential role and for this reason several interference reduction techniques are investigated. Special attention is also given to the case of low signal to noise ratio (SNR) and a way to evaluate the upper bound on EE in this regime is provided. This methodology leads to tractable analytical results under certain common channel properties, and thus allows the study of various models without the need for demanding system-level simulations.
\end{abstract}
\begin{IEEEkeywords}
Green wireless networks, spectral and energy efficiency, power and bandwidth allocation, stochastic geometry.
\end{IEEEkeywords}
\section{Introduction}
\IEEEPARstart{T}{he explosion} of data traffic in wireless networks in recent years with billions of daily mobile users, along with the corresponding exponential growth in infrastructure, has led to the rapid increase in the energy consumed by wireless networks. Technological innovations that solely allow the system components to consume less power clearly cannot keep pace with these changes. In this direction, although today's systems are mainly designed for optimal capacity with high target values for system throughput and spectral efficiency (SE), the operators are now showing greater interest in improving their energy efficiency (EE). The motivation behind this transition is the need to limit the electricity costs that represent a large portion of their operational expenditure (OPEX). Furthermore, from an equally important perspective, the rising concerns about global warming and environmental protection motivate the design of systems with improved EE and lower greenhouse gas emissions. This paradigm shift could directly benefit the EE of other energy-intensive sectors as well, through so-called smart technologies, including for instance smart energy grids, buildings and transportation control.

\IEEEpubidadjcol
\bstctlcite{IEEEbibChanges:BSTcontrol}

For all these reasons, a holistic approach for green wireless networks is widely envisaged and several significant actions in both academia and industry are already committed to this goal. For instance, the EARTH FP7 project investigated the development of a new energy efficient wireless generation \cite{EARTH} and the TREND FP7 Network of Excellence aimed to establish the integration of the European research community in green networking with a long term perspective \cite{TREND}. The Next Generation Mobile Networks (NGMN) Alliance \cite{NGMN} brought together partners with green activities and more recently, the ambitious mission of the GreenTouch Initiative is to deliver the roadmap in order to increase total network EE by a factor of 1000 compared to 2010 levels \cite{GreenTouch}.

However, the EE improvement is hardly a straightforward process when it is mainly based on green network management rather than technology and hardware advancements, as it can actually have a negative impact on other key performance indicators and particularly on SE. The optimization of both these metrics can lead to a challenging trade-off, since they are usually inter-related and conflicting. A fundamental insight is known from the well known Shannon formula for channel capacity with additive white Gaussian noise (AWGN), which shows that for a given data rate, transmitting with larger bandwidth leads to higher EE. The study of this trade-off is demanding for multi-user communications and becomes even harder in cellular networks. 

An interesting investigation of this issue, along with other fundamental ones about green wireless networks is described in \cite{GreenRadioTechniquesEdinburgh,YChenTradeoff,GTTCommMagazine,EEsurveyMiao}. Representative results on the SE-EE trade-off for single cell Orthogonal Frequency-Division Multiple Access (OFDMA) networks can also be found in \cite{YanEEdownlinkOFDMA,EEofdmaMiao,YanJointP-BW,akbariEEpimrc}, mainly aiming at algorithms for optimal resource block allocation, while an overview of game-theoretic approaches is presented in \cite{MeshkatiPoorEEinUplinkGames}. The extension to a multi-cell setup has so far not been extensively studied and to the best of our knowledge the respective publications are relatively limited. For instance, more general studies focusing on the cell level SE for cellular networks are described in \cite{RichterDeployment} and \cite{KarraySE-EE}, while an analysis for interference-limited scenarios is presented in \cite{TradeOffInterfLimited} and a simple one-dimensional (1-D) multi-cell environment is analyzed in \cite{GaultTradeoff} by introducing an asymptotic regime as the number of users grows to infinity. Other studies investigate more advanced architectures like distributed antenna systems (DAS) \cite{EEinDAS,EEVirtualMIMO}, multi-hop wireless networks \cite{TradeOffMultihop,JMGmultihop} or cognitive radio \cite{GraceCognitive}. Most of these works require complex system level simulations and are limited to scenarios with a fixed number of users. While simulations are necessary in order to evaluate in detail system performance, more tractable results are often desirable to easily reveal useful insights. Tools from stochastic geometry have been used in recent studies for this reason, as for example in \cite{MariosEEHetNet,ThroughputAndEEAnalysis,StochasticIA,MariosEE2Tier}, but these works significantly differ from our approach where the focus in on the SE-EE trade-off.

Along this line of thought, this paper presents a simple and practical theoretical framework for the analysis of the SE-EE relationship in cellular networks. To this end, a joint power-bandwidth allocation scheme for the downlink is adopted initially for a single cell multi-user scenario, under the assumption that only a statistical knowledge of the channel is available at the base station (BS).
A low complexity numerical solution is achieved and at the same time we prove that the cell traffic can be segmented into specific groups, allowing the study of various models and traffic distributions. Moreover, the introduced  model can take into account both transmit and signal processing power. Then, the special case of the low signal to noise ratio (SNR) regime is presented, where an explicit theoretical EE upper bound is defined. An extension of this framework to a multi-cell scenario is performed by assuming that the random locations of BSs form a Poisson point process (PPP). By applying properties of stochastic geometry, key metrics such as the interference and the signal to noise plus interference ratio (SINR) are analyzed, leading to a formulation similar to the one in the single cell case. Since interference plays a major role in this scenario, frequency reuse and beamforming are studied as potential interference reduction techniques, but our model broadly applies to many other approaches. 
Finally, it should be highlighted that the goal of this paper is to provide a way to easily obtain SE-EE trade-off curves and thus, we do not emphasize the comparison to other existing resource schemes. Interested readers are encouraged to refer to our results in \cite{JMGpimrc}. We summarize the key contributions of this work in the following points:
\begin{enumerate}
\item A simple approach for studying the SE-EE trade-off in the downlink of a single cell, with the help of a novel joint power-bandwidth allocation scheme.
\item A traffic repartition scheme that further reduces the complexity of the previous problem.
\item An extension to a multi-cell scenario with a PPP, where the SE-EE trade-off is still tractable.
\end{enumerate}

The remainder of the paper is organized as follows: Section II describes the single cell model, including the formulation of the optimization problem and the approach of traffic repartition. In Section III, the case of low SNR is presented and a representative example with uniform traffic is studied. The multi-cell scenario is discussed in Section IV and then, extensive numerical results are presented in Section V. Finally, our concluding remarks are made in Section VI.
\section{single cell model}

\subsection{System Model}
A single cell scenario is considered here, where our interest is focused on the downlink direction. The BS, located at the center of the cell, is assumed to serve a set $\mathcal{U}$ of randomly distributed users of cardinality $N_U$, while each user $u \in \mathcal{U}$ has a specific data rate demand $T_u$. The total available transmit power $P_{tot}$ and bandwidth $W_{tot}$ are shared among the $N_U$ users according to the applied resource allocation policy. The case of flat fading is addressed as a first step, but a similar approach can be followed even in the more complex case of frequency-selective fading, for example per OFDMA symbol, under certain conditions for the user channels \cite{GaultTradeoff}. In addition, since the formulation is based on the outage capacity, the channels are also assumed to be slowly-varying \cite{TseWirelessCommunBook}. Orthogonal multiple access in an AWGN channel is considered and for simplicity we neglect the intra-cell interference. Thus, for a random channel realization the achieved user capacity $C_u$ is given by the Shannon formula for an AWGN channel 
\begin{equation}
C_u = w_u \log_2 \left(1+\frac{1}{\gamma _\textit{eff}} \textit{SNR}_u\right)
\label{eq:C_u}
\end{equation}
where $w_u$ is the bandwidth allocated to user $u$ and $\gamma _\textit{eff}$ is the SNR gap that introduces the impact of practical modulation and coding schemes. Moreover, the SNR level of the user can be described in more detail by
\begin{equation}
\textit{SNR}_u = \frac{p_u h_u\ell_u}{w_u N_0}
\label{eq:SNR_u}
\end{equation}
where $p_u$ is the BS dedicated link transmit power, $N_0$ is the noise power spectral density, $h_u$ is the random variable that incorporates the effect of fading and finally, $\ell_u$ represents the deterministic part of the signal attenuation in the form of a proper path loss function.

\subsection{Performance Metrics}
Since we aim to study the SE-EE trade-off for different operational scenarios, the definitions of these key system performance indicators are briefly reviewed in line with the adopted system model.
\vspace{7pt}
\begin{definition}
The \textit{spectral} or \textit{bandwidth efficiency} is a measure that reflects the efficient utilization of the available spectrum in terms of throughput and it is commonly defined as the amount of throughput that the BS can transmit over a given bandwidth, expressed in $\text{bps}/\text{Hz}$. 
\end{definition}
\vspace{7pt} 

Hence, according to the definition of the outage capacity, by taking into account the probability that the channel gain is strong enough to support the traffic demand, the SE is given by
\begin{equation}
\textit{SE} = \frac{ \sum_{u \in \mathcal{U}} T_u \mathbb{P}\left[C_u \geq T_u\right]}{ \sum_{u \in \mathcal{U}} w_u}\, .
\label{eq:SE}
\end{equation}
\begin{definition}
The \textit{energy efficiency} on the other hand reflects the data transmission efficiency in terms of power consumption and it is defined as the amount of throughput that the BS can transmit per unit of power, expressed in $\text{bps}/\text{W}$ or $\text{bits}/\text{Joule}$. 
\end{definition}\vspace{7pt}

It is worth mentioning that in some scenarios, especially when coverage issues are studied, the area power consumption, expressed in $\text{W}/\text{m}^2$, is also practical as an alternative EE metric. In this work, the commonly used throughput-oriented notion is adopted which similarly to \eqref{eq:SE} leads to
\begin{equation}
\textit{EE} = \frac{ \sum_{u \in \mathcal{U}}T_u \mathbb{P}\left[C_u \geq T_u\right]}{\sum_{u \in \mathcal{U}} p_u} \, .
\label{eq:EE}
\end{equation}

Both efficiency metrics are defined so far according to the discrete set of cell users. These definitions can also be extended to the case of continuous traffic distributions over the cell coverage area by using surface integrals in \eqref{eq:SE}-\eqref{eq:EE}.

\subsection{Optimization Problem}
\label{sec: single cell opt problem}
Our objective is to maximize both efficiencies and at the same time satisfy the traffic demands, by properly allocating the BS resources among the users, i.e. both bandwidth $w_u$ and transmit power $p_u$ in this model. As a nontrivial multi-objective optimization problem, a single solution does not exist if no preference between the metrics is considered and for this reason we choose to provide the Pareto front of the respective trade-off curve. Since the resources are limited, the problem is subject to the following constraints: 
\begin{IEEEeqnarray}{rCl}
\sum\limits_{u \in \mathcal{U}} p_u & \leq & P_{tot} \IEEEyesnumber\IEEEyessubnumber \label{eq:InCondP}\\
\sum\limits_{u \in \mathcal{U}} w_u & \leq & W_{tot} \, . \IEEEyessubnumber \label{eq:InCondW} 
\end{IEEEeqnarray}

The probability in the SE-EE definitions can be analyzed by replacing $C_u$ according to \eqref{eq:C_u} and \eqref{eq:SNR_u}, which yields 
\begin{equation}
	\mathbb{P}\left[C_u \geq T_u\right]= \mathbb{P}\left[h_u\geq\frac{\gamma _\textit{eff}N_0w_u}{p_u\ell_u}\left(2^\frac{T_u}{w_u}-1\right)\right]  .
\label{eq:Pr_C>T_singlecell}
\end{equation}

By setting a threshold $0<c<1$ for this probability, 
in the general case where the fading follows an arbitrary distribution, \eqref{eq:Pr_C>T_singlecell} leads to the minimum required power
\begin{equation}
	p_u = \displaystyle \frac{\gamma _\textit{eff}N_0w_u }{\ell_u F_h^{-1}(1-c)}\left(2^\frac{T_u}{w_u}-1\right)
\label{eq:p_u_GeneralFading}
\end{equation}
where $F_h^{-1}(.)$ is the inverse cumulative distribution function (cdf) of $h$ whose index $u$ is omitted, since the fading distribution is assumed to be the same for all users.
\vspace{7pt}
\begin{remark}
In the typical case of Rayleigh fading, $h$ follows an exponential distribution and therefore for a mean value $\mathbb{E}\left[h\right]=1/\tau$, we obtain $F_h^{-1}(1-c)=\frac{1}{\tau}\ln\frac{1}{c}$.
\end{remark} \vspace{7pt}

Then, according to \eqref{eq:SE} and since both demand $T_u$ and threshold $c$ are specified, the SE becomes fixed for a given value of the total allocated bandwidth $W=\sum_u w_u\leq W_{tot}$. This remark allows us to easily find the set of Pareto optimal solutions by moving the effective trade-off point along the SE values. More precisely, the problem is equivalent to the one we obtain by maximizing EE for values of $W$ within the interval $\left(0,W_{tot} \right]$. Obviously, there are uncountably many real numbers inside any given interval and therefore we limit our analysis to a sufficient number of points for $W$, and SE respectively, that capture the Pareto front: \vspace{7pt} 

\noindent
\begin{tabular}{p{\linewidth-2\tabcolsep}}
\hline
	\begin{problem}
	\textit{SE -- EE optimization}
	\label{pr:singleCell}
	\end{problem}
	\vspace{-7pt}
	\\
	\hline
	\vspace{-5pt}
	\begin{equation*}
	\vspace{-5pt}
	(P_1) \quad
	\begin{array}{rl}
		\multicolumn{2}{l}{\underset{\left(p_u,w_u\right)}{\max} \textit{EE}, \quad \forall~ W\in \left(0,W_{tot} \right]}\\[7pt]
		\textit{s.t. }1. & p_u = \displaystyle \frac{\gamma _\textit{eff}N_0w_u }{\ell_u F_h^{-1}(1-c)}\left(2^\frac{T_u}{w_u}-1\right)\\[10pt]
		2. & \displaystyle  \sum_{u \in \mathcal{U} } w_u = W
	\end{array} 
	\end{equation*}\\
\hline
\end{tabular} \vspace{3pt}

Notice that if the solution of $(P_{\ref{pr:singleCell}})$ does not satisfy \eqref{eq:InCondP}, then there is no feasible solution and hence, this constraint is implicitly included. Several interesting conclusions can be derived from the theoretical analysis of $(P_{\ref{pr:singleCell}})$. Our first result is stated here, from which all the subsequent ones follow. \vspace{7pt}

\begin{theorem}
Given the description of $(P_{\ref{pr:singleCell}})$, the optimal allocation $\left(w_u^\textit{opt},p_u^\textit{opt}\right)$ to a user $u$ is found for bandwidth: 
\begin{equation}
	w_u^\textit{opt}=\dfrac{T_u\ln 2}{1+W_0\left(v\right)}, \text{with} ~ v \triangleq \frac{1}{e}\left\lbrace\frac{\lambda\ell_u F_h^{-1}(1-c)}{\gamma _\textit{eff}N_0}-1\right\rbrace
	\label{eq:w_u}
\end{equation}
and power:
\begin{equation}
	p_u^\textit{opt} = \frac{\gamma _\textit{eff} N_0 T_u\ln 2 }{\ell_u F_h^{-1}(1-c)} \cdot \frac{e^{1+W_0\left(v\right)}-1}{1+W_0\left(v\right)}
\label{eq:p_u}
\end{equation}
where $W_0$ is the principal branch of the real-valued Lambert function and $\lambda$ is the multiplier of the Lagrange function $\Lambda$:  
\begin{equation}
\Lambda\left(w_u,\lambda\right) = \displaystyle \sum\limits_{u \in \mathcal{U}} p_u + \lambda\left(\sum\limits_{u \in \mathcal{U}}w_u-W\right)  .
\label{eq:Lambda}
\end{equation}
\label{th:single}
\end{theorem}

\begin{IEEEproof}
Since the numerator in \eqref{eq:EE} is fixed for a specified threshold $c$, in order to maximize EE it is sufficient to minimize the allocated power while satisfying the problem constraints. Hence, a practical method for finding the solution of $(P_{\ref{pr:singleCell}})$ is to introduce the Lagrange multiplier $\lambda$, while the Lagrange function $\Lambda$ that we want to minimize is then the one defined in \eqref{eq:Lambda}. The solution should be a stationary point of $\Lambda$, where its partial derivatives are equal to zero, as follows:
\begin{equation}
\frac{\partial \Lambda}{\partial w_u}=0,~\frac{\partial \Lambda}{\partial \lambda}=0 \, .
\label{eq:LamdaDeriv}
\end{equation}

The first condition of \eqref{eq:LamdaDeriv} yields
\begin{equation}
		2^\frac{T_u}{w_u} \left(\dfrac{T_u \ln 2}{w_u}-1\right)= \frac{\lambda\ell_u F_h^{-1}(1-c)}{\gamma _\textit{eff}N_0}-1
	\label{eq:Lamda_deriv_w}
\end{equation}
and with the help of the equivalence $y=xe^x \Leftrightarrow x=W_0(y)$ that holds for the function $W_0$, we reach the expression of \eqref{eq:w_u}. Note that $x>-1$ in this case and only the single-valued principal branch of the Lambert function is used.  Then, substituting the value of $w_u^\textit{opt}$ into \eqref{eq:p_u_GeneralFading} leads to \eqref{eq:p_u}. It is also straightforward to see that the function $\Lambda$ is convex by applying the second derivative test in \eqref{eq:Lambda} and thus, the only critical point we get from \eqref{eq:w_u} leads by definition to the minimum required power and the desired maximum EE.
\end{IEEEproof} \vspace{3pt}

In order to find the exact solution of $(P_{\ref{pr:singleCell}})$, the value $\lambda$ must also be defined. A closed-form solution does not exist, but $\lambda$ can easily be computed numerically according to:\vspace{7pt}

\begin{lemma}
The Lagrange multiplier $\lambda$ that satisfies \eqref{eq:LamdaDeriv} for $(P_{\ref{pr:singleCell}})$ lies within a closed interval and is the root of a monotonic function. 
\label{lem:solution}
\end{lemma}  \vspace{3pt}

\begin{IEEEproof}
See Appendix \ref{app:lemma}.
\end{IEEEproof} \vspace{3pt}

Therefore, a root-finding algorithm such as the bisection method or more sophisticated ones can be applied, and then the optimal resource allocation is given by \eqref{eq:w_u} and \eqref{eq:p_u}.

\subsection{Traffic Repartition}
\begin{figure}[!t]
\centering
\includegraphics[width=0.9\linewidth]{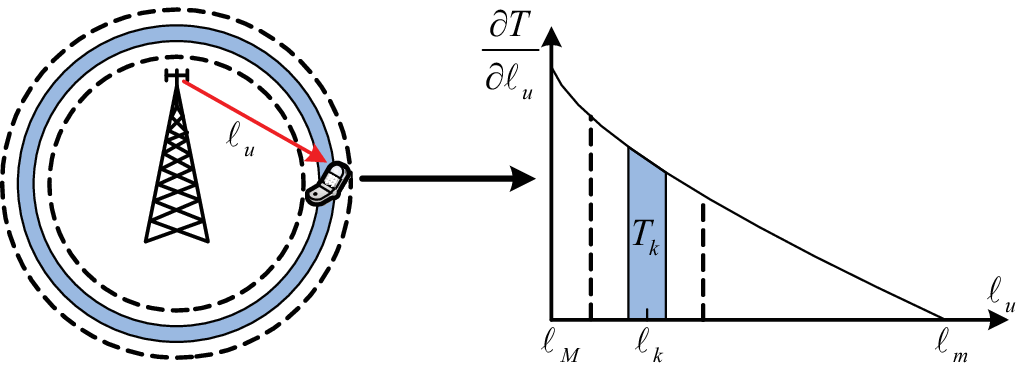}
\caption{Traffic repartition for a simple path loss model on the left figure; example of traffic volume as a function of attenuation on the right one.}
\label{fig:traffic} \vspace{-5pt}
\end{figure}

It is clear so far that in order to find the solution of $(P_{\ref{pr:singleCell}})$, the continuous variables $w_u$ and $p_u$ need to be calculated for all the cell users. However, the complexity can be reduced significantly with the help of \textit{Theorem} \ref{th:single}, by grouping together users who experience similar attenuation $\ell_u$. Specifically, the total cell traffic can be seen as a set $\mathcal{K}$ of $N_K$ partitions, each one characterized by a central attenuation $\ell_k$, with $k \in \mathcal{K}$. Then, the respective set of users $\mathcal{U}_k$ is formed by
\begin{equation}
u \in \mathcal{U}_k \Leftrightarrow \left|\ell_u-\ell_k\right| \leq \varepsilon_\ell
\label{eq:groups}
\end{equation}
where $\varepsilon_\ell$ is the interval size around $\ell_k$. Notice that besides equally spaced intervals $\varepsilon_\ell$, alternative spacings can be used to lead to more balanced traffic groups depending on the traffic distribution. Furthermore, the total cell traffic $T_{tot}$ is
\begin{equation}
T_{tot} \triangleq \displaystyle \sum\limits_{u \in \mathcal{U}}T_u = \sum\limits_{k \in \mathcal{K}} \sum\limits_{u \in \mathcal{U}_k}T_u = \sum\limits_{k \in \mathcal{K}}T_k
\label{eq:T_tot}
\end{equation}
where $T_k$ denotes the aggregated traffic demand from users of group $k$. It is worth mentioning that this way the study of continuous traffic distributions is also simplified. An example of this traffic repartition is illustrated on the left side of \figurename{\ref{fig:traffic}} for the simple case of a path loss model that depends only on the distance between the user and the BS. The shaded area represents the traffic partition of a specific group $k$. The right plot further assumes a continuous uniform traffic distribution $T$ and shows the volume of this traffic as a function of the attenuation. In this figure, $\ell_{m}$ and $\ell_{M}$ refer respectively to the minimum and maximum attenuation for users close to the BS and at the cell edge respectively, with $\ell_{m} > \ell_{M}$. 

Conveniently, this approach allows us to allocate to each set of users $\mathcal{U}_k$ an overall bandwidth $W_k=\sum_{u \in \mathcal{U}_k}w_u$ and transmit power $P_k=\sum_{u \in \mathcal{U}_k}p_u$ in a similar way to $(P_{\ref{pr:singleCell}})$ and specifically by using \eqref{eq:w_u} and \eqref{eq:p_u} of \textit{Theorem} \ref{th:single}. The new problem is presented here for the sake of completeness. \vspace{7pt}

\noindent
\begin{tabular}{p{\linewidth-2\tabcolsep}}
\hline
	\begin{problem}
	\textit{SE -- EE optimization with traffic repartition}
	\label{pr:repartition}
	\end{problem}
	\vspace{-7pt}
	\\
	\hline
	\vspace{-5pt}
	\begin{equation*}
	\vspace{-5pt}
	(P_2) \quad
	\begin{array}{rl}
		\multicolumn{2}{l}{\hspace*{-20pt} \underset{\left(P_k,W_k\right)}{\max} \textit{EE}, ~\forall~ W\in \left(0,W_{tot} \right],~u \in \mathcal{U}_k \Leftrightarrow \left|\ell_u-\ell_k\right| \leq \varepsilon_\ell}\\ [10pt]
		\textit{s.t. }1. & P_k = \displaystyle \frac{\gamma _\textit{eff}N_0W_k }{\ell_u F_h^{-1}(1-c)}\left(2^\frac{T_k}{W_k}-1\right)\\[10pt]
		2. & \displaystyle  \sum_{k \in \mathcal{K}} W_k = W
	\end{array}
	\end{equation*}\\
\hline
\end{tabular} \vspace{3pt}

Besides this allocation scheme on a group level, a method to derive the respective values for each individual user is still needed. This proves to be quite simple, as stated in the next theorem, and hence, justifies the traffic repartition approach.

\begin{theorem}
The optimal resource allocation $\left(w_u^*,p_u^*\right)$ for a user $u \in \mathcal{U}_k$, given the solution $(W_k,P_k)$ of $(P_{\ref{pr:repartition}})$, is achieved for both bandwidth and transmit power proportional to traffic $T_u$, and specifically for $w_u^*=\frac{T_u}{T_k}W_k$ and $p_u^*=\frac{T_u}{T_k}P_k$. \vspace{7pt}
\label{th:repartition}
\end{theorem}

\begin{IEEEproof}
See Appendix \ref{app:th2}. 
\end{IEEEproof} \vspace{3pt}

\begin{remark}
With the help of \textit{Theorem} \ref{th:repartition}, the number of optimization variables is reduced from $2 \times N_U$ to $2 \times N_K$.
\end{remark} \vspace{7pt}

At this point, one should notice that since the users are allocated resources according to the central attenuation $\ell_k$ of their group and not their individual $\ell_u$, the achieved performance is expected to be lower compared to the optimal one without traffic repartition. However, as we illustrate through our numerical results, a small number $K$ of groups is sufficient to reach similar performance even for a practical case of a large cell, where the values of attenuation vary significantly within its coverage area. 

\subsection{Signal Processing Power}
The formulation of the problem is based until now on the evaluation of EE from \eqref{eq:EE}, where only the transmit power is considered. A more accurate model could also take into account other components that contribute to the overall BS power consumption. For instance, the signal processing power can play an important role due to various BS components in baseband processing and radio frequency circuits that actually consume more power as the allocated bandwidth $W$ increases. A simplified approach is to assume that this processing power is proportional to $W$ by a constant factor $P_{\textit{SP}}$, expressed in $\text{W}/\text{Hz}$, as proposed for example in \cite{YanJointP-BW}. As a consequence, a new problem can be formed by modifying properly the optimization function of $(P_{\ref{pr:singleCell}})$: \vspace{7pt} 

\noindent
\begin{tabular}{p{\linewidth-2\tabcolsep}}
\hline
	\begin{problem}
	\textit{SE -- EE optimization with processing power}
	\label{pr:P_SP}
	\end{problem}
	\vspace{-7pt}
	\\
	\hline
	\vspace{-3pt}
	\begin{equation*}
	\vspace{-3pt}
	\!(P_3)
	\begin{array}{rl}
		\multicolumn{2}{l}{\underset{(p_u,w_u)}{\max} \textit{EE}= \dfrac{\sum_{u \in \mathcal{U}}T_u \mathbb{P}\left(C_u \geq T_u\right)}{\sum_{u \in \mathcal{U}}p_u+P_{\textit{SP}}W}, \forall~ W \in \left(0,W_{tot} \right]}\\[15pt]
		\textit{s.t. } & \text{the constraints of } (P_{\ref{pr:singleCell}})
	\end{array}
	\end{equation*}\\
\hline
\end{tabular} \vspace{3pt}

The optimal solution in this case is the same as in $(P_{\ref{pr:singleCell}})$, as the additional term $P_{\textit{SP}}W$ that appears in \eqref{eq:Lambda} vanishes when the first condition of \eqref{eq:LamdaDeriv} is applied. The only difference comes from the evaluation of EE for each trade-off point along the SE values.  
Note that the same model can be applied to $(P_{\ref{pr:repartition}})$ as well. The formulation of $(P_{\ref{pr:P_SP}})$ represents a more general case compared to the previous problems that easily follow for $P_{SP}=0$. However, it is intentionally presented last since most of our results are related to the case in which only the transmit power is considered. Finally, as readers more familiar with power models may identify, another part of the BS power consumption breakdown could be included as well, describing the power figure that is independent of the bandwidth, e.g. as a result of the main power supply or the cooling system \cite{TsilimantosBCG}. This model can be addressed similarly to $(P_{\ref{pr:P_SP}})$.
\section{Low SNR Regime}
A special case of interest is presented here in order to find the theoretical EE upper bound without solving the previous optimization problems. The adopted process is not new, as it is quite standard how to simplify the problem in the wideband regime \cite{VerduWidebandRegime}. However, we briefly present the achieved bound in line with our formulation, as it complements the analysis.

\subsection{EE Upper Bound}
According to formula \eqref{eq:C_u} for channel capacity, as the allocated bandwidth increases, the capacity remains finite and reaches an asymptotic limit. This is the low SNR regime of the AWGN channel with only a power constraint and no limitation on bandwidth. The needed transmit power in this case is the lowest possible leading to the highest EE. This is described by the following lemma: \vspace{7pt}

\begin{lemma}
\label{lem:lowSNR}
The achieved optimal EE of $(P_{\ref{pr:singleCell}})$ admits the following explicit upper bound $\textit{EE}^*$ as $W\rightarrow \infty$: 
\begin{equation}
\textit{EE}^* = \frac{cT_{tot}F_h^{-1}(1-c)}{\gamma _\textit{eff}N_0 \ln 2  \sum\limits_{u \in \mathcal{U}}\frac{T_u}{\ell_u}}  \, .
\label{eq:EE_tot lowSNR}
\end{equation}
\end{lemma}

\begin{IEEEproof}
See Appendix \ref{app:lowSNR}.
\end{IEEEproof} \vspace{3pt}

This maximum value of EE is computed very easily if the attenuation values $\ell_u$ are known.
One should keep in mind that although operating in the low SNR regime is better in terms of EE, this approach brings also new challenges. Besides the fact that the bandwidth is in general limited, the design of appropriate receivers that perform sufficiently in this region can become a key issue, since in this case the performance of several tasks, such as carrier and clock recovery and frame synchronization, can decrease significantly.

\subsection{Simple Case Study with Uniform Traffic}
A simple theoretical model is studied here in order to derive more specific results for the EE bound of \eqref{eq:EE_tot lowSNR}. A single cell of radius $R_C$ is assumed for this purpose, along with a continuous uniform traffic distribution with constant density $T_0$ in $\text{bps}/\text{m}^2$. The attenuation is considered to be a result of a simple power law path loss model, described by
\begin{equation}
\ell_u=\left(\kappa d\right)^{-\alpha},\quad \text{with} ~ d=\sqrt{L^2+R^2}
\label{eq:PLmodel}
\end{equation}
where $\kappa>0$ is a scenario dependent constant, $\alpha>2$ is the path loss exponent, $L>\frac{1}{\kappa}$ is the BS antenna height and $R$ is the horizontal distance between the user $u$ and the BS. Notice that this model, as proposed in our previous work in \cite{TsilimantosBCG}, is actually a modified version of the commonly used power law model. Specifically, the parameter $L$ is introduced in order to avoid the singularity for $R=0$, which leads to invalid scenarios, since the received power exceeds the transmitted one as the mobile moves closer to a BS. An interesting study of the impact from unbounded path loss models on network performance can be found in \cite{ModifiedPL}. Finally, by using polar coordinates, \eqref{eq:EE_tot lowSNR} becomes
\begin{equation}
		\textit{EE}^* = \frac{c\left(\alpha+2\right) R^2_CF_h^{-1}(1-c)}{2\gamma _\textit{eff}N_0\kappa ^\alpha  \ln 2 \left[\left(L^2+R^2_C\right)^{\alpha/2+1}-L^{\alpha+2}\right]} \, .	
	\label{eq:EE_contLowSNR} 
\end{equation}

The EE upper bound in this case no longer depends on the traffic demand, but only on the cell size and the path loss model parameters. We will return to this expression in the section on numerical results.
\section{Multi-Cell Scenario}
It is well known that the impact of interference is crucial for the performance of wireless networks. This effect is neglected in the previous analysis where a single cell scenario is investigated and the user capacity is defined as a function of the achieved SNR. This section presents an extension of the theoretical approach, based on stochastic geometry properties, that allows us to characterize key system metrics, such as the SINR distribution, and eventually form an optimization problem similar to $(P_{\ref{pr:repartition}})$ to which our solution process still applies. To this end, firstly the multi-cell system model is described, following a similar process as in \cite{TsilimantosBCG} and \cite{BaccelliNOW,BaccelliCellular,BaccelliJSAC}.

\subsection{System Model}
A large-scale cellular network is considered here, where the random distribution of BSs forms a two-dimensional \text{(2-D)} homogeneous PPP $\Phi_\text{B}$ with intensity $\lambda_\text{B}$. This is a very useful choice among other stochastic models due to its well known properties and, from a practical point of view, it leads to a topology of various cell shapes and sizes that usually approximates the actual deployment in large wireless networks. Our interest is still focused on the downlink direction without intra-cell interference, where specifically the mobile users are assumed to be served by their nearest BS. Unlike the model in \cite{TsilimantosBCG}, where the resources are equally shared among the users, here the analysis of the optimal bandwidth and power allocation $\left(w_u,p_u\right)$ is performed for the set of users $\mathcal{U}_b$ of one randomly selected BS $b$. All the other BSs have total available bandwidth $W_I$ and transmit power $P_I$, serve at least one user and are assumed to be at full load with constant transmit power spectral density, i.e. the allocated transmit power $p_i$ from BS $i$ that interferes with a user $u \in \mathcal{U}_b$ is proportional to the bandwidth $w_u$ according to
\begin{equation}
p_i = w_u \frac{P_I}{W_I} ~ \forall i \in \left\lbrace \Phi_\text{B} \backslash b \right\rbrace  .
\label{eq:pInterf}
\end{equation}

\begin{figure}[!t]
\centering
\includegraphics[width=0.8\linewidth]{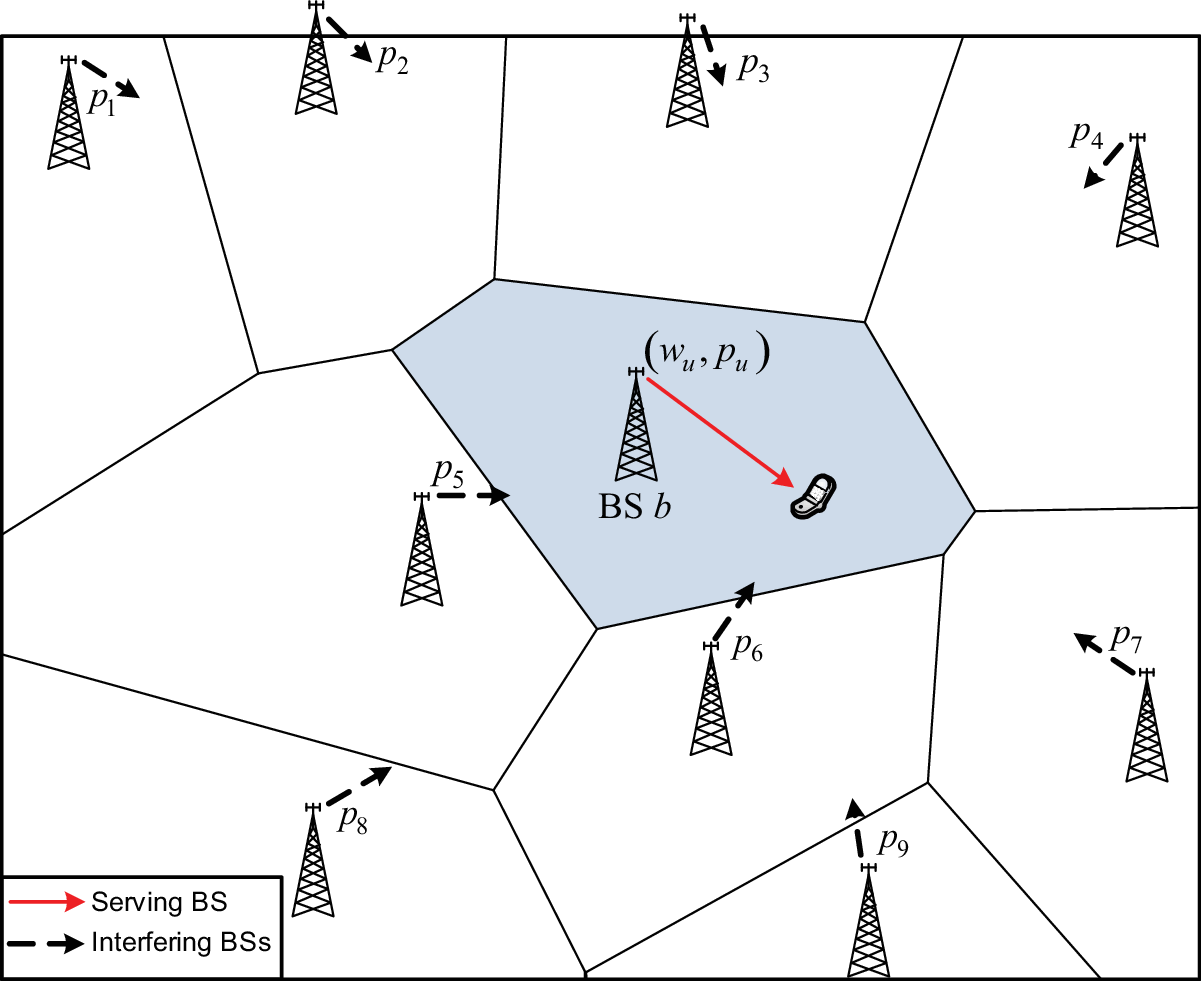}
\caption{Example of multi-cell topology with a PPP $\Phi_\text{B}$ of 10 cells. The shaded polygon shows the serving area of BS $b$. }
\label{fig:multicellModel} \vspace{-5pt}
\end{figure}

An example of this topology is shown in \figurename{\ref{fig:multicellModel}} for a random realization of $\Phi_\text{B}$. It should be mentioned that power control schemes can also be studied, as for example in the case of general fading for interfering signals. Further details on this follow in the respective part of this section. Thereby, unless otherwise stated, a scenario with full resource utilization is obtained, where each mobile in $\mathcal{U}_b$ receives interference from all the points of $\Phi_\text{B}$ besides the serving BS $b$. Even though this model is not a full multi-cell scenario, which is known to be hard to deal with, as shown for example in \cite{MultiCellGameTheory}, it still plays a significant role as it can serve as a worst case scenario in terms of interference. The gains from interference management and resource allocation for all the BSs is a subject for future study.

\subsection{SINR Analysis}
By using the Shannon capacity formula with interference treated as noise, the achieved user capacity of a user $u \in \mathcal{U}_b$ from \eqref{eq:C_u} can now be replaced by
\begin{equation}
C_u = w_u \log_2 \left(1+\frac{1}{\gamma_\textit{eff}} \textit{SINR}_u\right).
\label{eq:CuSINR}
\end{equation}

The generic expression for the SINR level measured at this user $u$ is given by
\begin{equation}
\textit{SINR}_u = \frac{S}{w_uN_0 + I}
\label{eq:SINRu}
\end{equation}
where $S$ is the desired dedicated signal and $I$ is the received interference. Both of these signals are subject to the stochastic nature of the PPP. Considering the attenuation as in the case of the single cell model, the signal $S$ can be described in more detail according to
\begin{equation}
S = p_u h_u \ell_u(R)
\label{eq:S}
\end{equation}
where $\ell_u$ is now written as a function of the distance $R$ between the user $u$ and the serving BS $b$. On the other hand, the received interference $I$ from all interfering BSs is 
\begin{IEEEeqnarray}{l}
I = \displaystyle \sum\limits_{i \in \left\lbrace \Phi_\text{B} \backslash b \right\rbrace} p_ig_{u,i}\ell_u (r_i)= w_u \frac{P_I}{W_I}\sum\limits_{i \in \left\lbrace \Phi_\text{B} \backslash b \right\rbrace}g_{u,i}\ell_u (r_i)
\IEEEeqnarraynumspace \label{eq:I}
\end{IEEEeqnarray}
with the right-hand equation simply derived by using \eqref{eq:pInterf}. The summation is performed over any interfering BS $i$ with respective distance $r_i$ from user $u$ and fading gain $g_{u,i}$. By substituting the last expressions into \eqref{eq:SINRu}, we obtain
\begin{equation}
\textit{SINR}_u = \frac{p_u}{w_u} \frac{h_u\ell_u(R)}{N_0+\frac{P_I}{W_I}\displaystyle \sum\limits_{i \in \left\lbrace \Phi_\text{B} \backslash b \right\rbrace}g_{u,i}\ell_u (r_i)}\triangleq \frac{p_u}{w_u}Y_u
\label{eq:SINRuFull}
\end{equation}
where $Y_u$ is a random variable with a distribution independent from the resource allocation $\left(w_u,p_u\right)$  and expressed in $\text{Hz}/\text{W}$. Conditioning on the distance $R$, the probability that the SINR is higher than a specific threshold $\rho$, i.e. the coverage probability, is then
\begin{equation}
\mathbb{P}\left[\textit{SINR}_u>\rho\middle|R\right]= \mathbb{P}\left[h_u>\frac{\rho\left(w_uN_0+I\right)}{p_u\ell_u (R)}\middle|R\right].
\label{eq:Pr_C}
\end{equation}

For a randomly located user, the cdf of the distance $R$ to the nearest BS, i.e. to the closest point of the PPP, is given by
$\mathbb{P}\left[R\leq x\right] =1 - e^{-\pi \lambda_{\text{B}}x^2}$ and as a result, the expression for the coverage probability $P_C$ averaged over the plane becomes
\begin{equation}
P_C(\rho)=2\pi\lambda_{\text{B}} \int\limits_{0}^{\infty}\mathbb{P}\left[\textit{SINR}_u>\rho\middle|R\right]e^{-\pi \lambda_{\text{B}}R^2}R\, \mathrm{d}R \, .
\label{eq:Pc_Final}
\end{equation}

At this point there are several different techniques that can be used to advance the analysis. An interesting survey of the related literature is presented in \cite{StochasticSurvey}. In the remainder of this section we focus on two of the most widely adopted approaches, the assumption of Rayleigh fading and the approximation of the interference by a known distribution.

\subsubsection{Rayleigh fading on the desired link}
This technique is extensively used in the literature due to its simplicity and analytical tractability. Specifically, for Rayleigh fading on the link of the desired signal $S$ with $\mathbb{E}\left[h\right]=1/\tau$, \eqref{eq:Pr_C} leads to
\begin{IEEEeqnarray}{l}
\mathbb{P}\left[\textit{SINR}_u>\rho\middle|R\right]=e^{-sw_uN_0}\mathcal{L}_I(s) \, , \displaystyle \text{ with}~s\triangleq\frac{\tau \rho}{p_u \ell_u (R)} \, . \IEEEeqnarraynumspace
\label{eq:Pr_C_Laplace}
\end{IEEEeqnarray}

Moreover, with the help of the probability generating functional (PGFL) \cite{StoyanStochastic}, the introduced Laplace transform of the interference $\mathcal{L}_I(s)\triangleq \mathbb{E}\left[e^{-sI}\right]$ is given by
\begin{IEEEeqnarray}{l}
\mathcal{L}_I(s)= \exp \left\lbrace -2\pi\lambda_{\text{B}} \int\limits_{R}^{\infty} \left(1-\mathbb{E}_g\left[e^{-spg\ell_u (r)} \right]\right) \mathrm{d}r \right\rbrace \IEEEeqnarraynumspace
\label{eq:Laplace}
\end{IEEEeqnarray}
where the expectation is performed over the fading gain of the interfering links whose index $u$ is dropped hereafter, since $g_{u,i}$ is assumed to follow the same distribution for all $(u,i)$ links. Notice also that the index $i$ for the discrete set of interferers is omitted in the case of the continuous variable $r$. Moreover, the integration limits are $\left[R,\infty\right)$, as the user is connected to the nearest BS $b$ and all the interfering BSs cannot be closer. 

One should note that the cdf of the variable $Y_u$ is equal to $F_Y(y)=\mathbb{P}\left[Y_u\leq y\right]=1-P_C\left(\frac{yp_u}{w_u}\right)$ according to \eqref{eq:SINRuFull} and \eqref{eq:Pc_Final}. Then, following the lines of \cite{BaccelliCellular} for the further analysis of $\mathcal{L}_I(s)$ and $P_C(\rho)$, we finally reach the expression
\begin{equation}
	\displaystyle F_Y(y)= 1-\pi \! \int\limits_{\lambda_{\text{B}}L^2}^{\infty}e^{\pi\lambda_{\text{B}}L^2\left[ J_1(x,y)+J_2(x,y)\right]}\, \mathrm{d}x \, .
\label{eq:Ydistr}
\end{equation}

In the case in which interference experiences general fading, the additional functions in \eqref{eq:Ydistr} are defined as
\begin{IEEEeqnarray}{c}
	J_1(x,y)= -y\tau  N_0\kappa^\alpha\left(\frac{x}{\lambda_{\text{B}}}\right)^{\frac{\alpha}{2}} \IEEEeqnarraynumspace \IEEEyesnumber\IEEEyessubnumber \label{eq:J1GeneralFading}\\
	J_2(x,y)= \frac{2\pi x}{\alpha} \left(\frac{y \tau P_I}{W_I} \right)^{\frac{2}{\alpha}} \! \mathbb{E}_g\!\left[g^{\frac{2}{\alpha}}\gamma \left(-\frac{2}{\alpha},\frac{y \tau g P_I}{W_I}\right)\right] \IEEEeqnarraynumspace \IEEEyessubnumber \label{eq:J2GeneralFading}
\end{IEEEeqnarray}
where the path loss model of \eqref{eq:PLmodel} is used, the interfering power $p$ and the argument $s$ are replaced from \eqref{eq:pInterf} and \eqref{eq:Pr_C_Laplace} respectively and $\gamma\left(s,x\right)=\int_{0}^{x}t^{s-1}e^{-t}\, \mathrm{d}t$ is the lower incomplete gamma function. This result can also include the scenario of power control schemes where $g$ is assumed to capture this variation, since from an interference point of view, it does not matter if the randomness is a result of the power $p_i$ or the fading $g$ or their product that appears in our model. Finally, in the special case where Rayleigh fading is considered for the interfering signals as well with the same mean value $\mathbb{E}\left[g\right]=1/\tau$, only \eqref{eq:J2GeneralFading} changes and becomes
\begin{equation}
	\displaystyle J_2^R(x,y)= -\pi x \left[1+I_\alpha\left(\frac{y P_I}{W_I}\right)\right], ~\text{with} 
\label{eq:J2Rayleigh}
\end{equation} \vspace{-15pt}
\begin{equation*}
\displaystyle I_\alpha(x)\triangleq \!\! \int\limits_{1}^{\infty}\frac{x}{x+z^{\alpha/2}} \, \mathrm{d}z = \frac{2x}{\alpha-2}~{}_2F_1\left(1,1-\frac{2}{\alpha};2-\frac{2}{\alpha};-x\right)
\end{equation*}
where ${}_2F_1$ is the Gaussian hyper-geometric function, available in many numerical computing packages (e.g. \verb|hypergeom| in Matlab). Finally, it should be mentioned that the results of this subsection for Rayleigh fading in the desired link are similar to the respective ones in \cite{BaccelliCellular}, with the few differences stemming from the adopted non-singular path loss model and the expression for the interfering power. 

\subsubsection{Approximation using the gamma distribution}
The second technique is based on the idea of approximating the aggregate interference by a known distribution. In our case, both the desired signal $S$ and the interference $I$ are considered to be drawn from a gamma distribution and second order moment matching is performed. Besides its simplicity in the achieved results, this technique allows the modeling of several fading distributions, e.g. Rayleigh and Nakagami fading, and different transmission modes with diversity, including maximum ratio combining and multiuser multiple-input multiple-output (MIMO). Interested readers can refer to \cite{HeathGammaDistribution} for further details. Our analysis follows the approach in \cite{HeathGammaDistribution}, but a different network topology is studied.

In order to begin the analysis, the fading $h$ is considered to follow a gamma distribution $\Gamma \left[k_h,\theta  _h\right]$, where $k_h$ and $\theta_h$ are the shape and scale parameters respectively. Then, due to the scaling property of the gamma distribution it is easily seen that $S\sim \Gamma \left[k_s,\theta_s\right]$, where $k_s=k_h$, $\theta_s=p_u\ell_u(R)\theta_h$ and $S$ is still given by \eqref{eq:S}. Regarding the interference of \eqref{eq:I}, it is further assumed that $I\sim \Gamma \left[k_I,\theta_I\right]$, where the distribution parameters are found below. Notice that the interference moments exist since the non-singular path loss model of \eqref{eq:PLmodel} is used. By using Campbell's theorem, the first moment of $I$ is 
\begin{equation}
	\displaystyle \mathbb{E}[I]= 2\pi\lambda_{\text{B}}\mathbb{E}\!\left[g\right]\int\limits_{R}^{\infty}p\ell_u(r)r\, \mathrm{d}r  = \frac{2\pi\lambda_{\text{B}}p\mathbb{E}\!\left[g\right]d^{2-\alpha}}{\kappa^\alpha \left(\alpha-2\right)} 
\label{eq:FirstMomentInterference}
\end{equation}
and the second moment is
\begin{IEEEeqnarray}{l}
	\displaystyle \mathbb{V}[I]= 2\pi\lambda_{\text{B}}\mathbb{E}\!\left[g^2\right]\!\int\limits_{R}^{\infty}\!p^2\ell_u^2(r)r\, \mathrm{d}r  = \frac{\pi\lambda_{\text{B}}p^2\mathbb{E}\!\left[g^2\right]\!d^{2-2\alpha}}{\kappa^{2\alpha} \left(\alpha-1\right)}\, . \IEEEeqnarraynumspace
\label{eq:SecondMomentInterference}
\end{IEEEeqnarray}

Then, according to second order moment matching, the parameters $k_I$ and $\theta_I$ are equal to
\begin{IEEEeqnarray}{rCl}
	\displaystyle k_I = \frac{\mathbb{E}^2\left[I\right]}{\mathbb{V}\left[I\right]}&=&\frac{4\pi \lambda_{\text{B}}d^2 \left(\alpha-1\right) }{\left(\alpha-2\right)^2}\frac{\mathbb{E}^2\left[g\right]}{\mathbb{E}\left[g^2\right]}  \IEEEyesnumber\IEEEyessubnumber \\ [5pt]
	\displaystyle \theta_I = \frac{\mathbb{V}\left[I\right]}{\mathbb{E}\left[I\right]}&=&\frac{p\left(\alpha-2\right) }{2\kappa^\alpha d^\alpha \left(\alpha-1\right)}\frac{\mathbb{E}\left[g^2\right]}{\mathbb{E}\left[g\right]} \, .  \IEEEyessubnumber 
\label{eq:MomentMatching}
\end{IEEEeqnarray}

Moreover, the coverage probability is now expressed as
\begin{IEEEeqnarray}{rCl}
	\mathbb{P}\left[\textit{SINR}_u>\rho\middle|R\right] &=& 1-\mathbb{E}_I\left[F_S(\rho\left(I+w_uN_0\right))\right] \IEEEeqnarraynumspace  \nonumber\\
	&=& \displaystyle 1-\int\limits_{0}^{\infty}F_S(\rho\left(x+w_uN_0\right))f_I(x) \, \mathrm{d}x \IEEEeqnarraynumspace
\label{eq:Pr_C_Gamma}
\end{IEEEeqnarray}
where $F_S$ is the cdf of $S$ and $f_I$ is the probability density function (pdf) of $I$. Both these functions are known for the gamma distribution and specifically
\begin{equation}
	\displaystyle F_S(x)=\frac{\gamma\left(k_s,x/\theta_s\right)}{\Gamma(k_s)}, ~ f_I(x)=\frac{x^{k_I-1}e^{-x/\theta_I}}{\theta_I^k ~ \Gamma(k_I)}
\label{eq:GammaCDF_PDF}
\end{equation}
with $\Gamma(t)=\int_{0}^{\infty}x^{t-1}e^{-x} \, \mathrm{d}x$ denoting the gamma function. A simplification that leads to more tractable results is to neglect the noise \cite{HeathGammaDistribution}, i.e. $\textit{SINR}_u\rightarrow\textit{SIR}_u=S/I$ that leads to
\begin{IEEEeqnarray}{rCl}
		\displaystyle \mathbb{P}\left[\textit{SIR}_u>\rho\middle|R\right] &=& \frac{\Gamma(k_I+k_s)}{\Gamma(k_s)\Gamma(k_I+1)}\left(\frac{\theta_s}{\rho\theta_I}\right)^{k_I} \nonumber \IEEEeqnarraynumspace \\
	  &\cdot& {}_2F_1\left(k_I,k_I+k_s;k_I+1;-\frac{\theta_s}{\rho \theta_I}\right). \IEEEeqnarraynumspace
\label{eq:P_SIR_Gamma}
\end{IEEEeqnarray}

Therefore, combining \eqref{eq:Pc_Final} with \eqref{eq:Pr_C_Gamma} or \eqref{eq:P_SIR_Gamma}, and replacing the values of $k_s,\theta_s,k_I$ and $\theta_I$, the cdf of the variable $Y_u$ can still be found as $F_Y(y)=1-P_C\left(\frac{yp_u}{w_u}\right)$. As a final remark on the gamma approximation technique, it should be mentioned that although it is very helpful when the Rayleigh fading approach is not appropriate, its main drawback is that simulations are needed to validate the achieved approximation.

\subsection{Optimization Problem}
Given the cdf $F_Y(y)$, it is now possible to follow a similar approach as in $(P_{\ref{pr:repartition}})$ with $N_K$ groups of cellular traffic. The difference here is that each group $\mathcal{U}_k$ is characterized by a central value of $Y_u$, denoted by $y_k$, instead of the attenuation $\ell_k$. The corresponding optimization problem for an interval size $\varepsilon_y$ around $y_k$ is presented below:\vspace{7pt} 

\noindent
\begin{tabular}{p{\linewidth-2\tabcolsep}}
\hline
	\begin{problem}
	\textit{SE -- EE optimization in the multi-cell scenario}
	\label{pr:multicell}
	\end{problem}
	\vspace{-15pt}
	\\
	\hline
	\vspace{-5pt}
	\begin{equation*}
	\vspace{-5pt}
	(P_4) \quad
	\begin{array}{rl}
		\multicolumn{2}{l}{\hspace*{-22pt} \underset{\left(P_k,W_k\right)}{\max} \textit{EE}, ~\forall~ W\in \left(0,W_{tot} \right],~u \in \mathcal{U}_k \Leftrightarrow \left|Y_u-y_k\right| \leq \varepsilon_y}\\[10pt]
		\textit{s.t. }1. & P_k = \dfrac{\gamma_\textit{eff} W_k}{y_k}\left(2^\frac{T_k}{W_k}-1\right)\\ [10pt]
		2. & \displaystyle  \sum_{k \in \mathcal{K}} W_k = W
	\end{array}
	\end{equation*}\\
\hline
\end{tabular} \vspace{3pt}

From this point, the optimal allocation $\left(W_k,P_k\right)$ is found by following the process of solving $(P_{\ref{pr:repartition}})$. Even though the provided SE-EE trade-off is based on average values due to the random nature of the PPP, the respective numerical results can be very useful in order to additionally assess the impact of interference and evaluate the gains from possible interference mitigation techniques. One should also notice that in this case, instead of setting a value for the probability $\mathbb{P}\left(C_u \geq T_u\right)$ as for the single cell scenario, the users with SINR below a specific threshold are assumed to be in outage.\vspace{7pt}

\begin{remark}
\label{rm:multicell}
For uniform traffic density $T_0$, the respective mean cell traffic for BS $b$ is equal to $\frac{T_0}{\lambda_{\text{B}}}$, and the traffic demand for each group $T_k$ can be found from the distribution of $Y_u$ as follows: 
\begin{IEEEeqnarray}{rCl}
	T_k &=& \dfrac{T_0}{\lambda_{\text{B}}} \mathbb{P}\left[\left|Y_u-y_k\right| \leq \varepsilon_y\right] \nonumber\\[5pt]
	&=&\dfrac{T_0}{\lambda_{\text{B}}}\left[ F_Y(y_k+\varepsilon_y)-F_Y(y_k-\varepsilon_y) \right] .
\label{eq:T_k_Multicell}
\end{IEEEeqnarray}
\end{remark}
 
\subsection{Interference Reduction Techniques}
The presented multi-cell theoretical analysis allows us to examine interesting techniques for interference reduction and performance improvement by slightly modifying the adopted mathematical model. In this context, the impact of frequency reuse and beamforming are investigated in this section. For simplicity, we limit our study to the first technique with Rayleigh fading on both the desired link and on the interfering signals.
\vspace{3pt}
\subsubsection{Frequency Reuse}
A typical way to reduce interference in wireless networks is to control the number of interfering BSs. In OFDMA based systems such as Long Term Evolution (LTE), it is possible to configure several different frequency patterns \cite{LTE_HolmaToskala}. A simple frequency reuse pattern is studied here by introducing a factor $f_r \geq 1$ for the number of BSs that can use the same frequency for transmission. Specifically, each cell can allocate a set of frequency channels that correspond to a total bandwidth $W/f_r$. In the case of a PPP, this procedure is described by the PPP thinning property that leads to a new PPP with intensity $\lambda_\text{B}/f_r$. Thus, by replacing the intensity of the interfering BSs and adjusting the available BS bandwidth, \eqref{eq:J1GeneralFading} remains the same while \eqref{eq:J2GeneralFading} becomes
\begin{equation}
	\displaystyle J_2^{F}(x,y)= -\pi x \left[1+\frac{1}{f_r}I_\alpha\left(\frac{y f_r P_I}{W_I}\right)\right]
\label{eq:FreqReuse}
\end{equation}
and the second constraint of $(P_{\ref{pr:multicell}})$ is now $\sum\limits_{k \in \mathcal{K}}W_K=W/f_r$.
\vspace{7pt}
\subsubsection{Beamforming}
Another approach towards interference mitigation is the utilization of advanced antenna techniques. The conventional delay and sum beamformer is considered as a first step in our model, where the BSs are equipped with a uniform linear array of $N_t$ antenna elements and spacing of half wavelength. For simplicity, the array axis is set in a way that a served mobile is always in boresight direction, as in \cite{DecreusefondBeamFormer}. In order to have a fair comparison, the gain of a single antenna element is set equal to one. Hence, the radiation pattern of the beamformer is given by
\begin{equation}
	A(\theta) = \left\{
	\begin{array}{cl}
		\dfrac{1}{N_t}\left| \dfrac{\sin \left(\frac{N_t \pi}{2} \cos \theta\right)}{\sin \left(\frac{\pi}{2} \cos \theta\right)} \right|^2  & \text{if } 0 \leq \theta \leq \pi,\\[15pt]
		-G_\textit{FB} & \text{else }
	\end{array} \right.
	\label{eq:RadiationPattern}
\end{equation}
for an arbitrary azimuthal direction $\theta$ measured with respect to array axis, while $G_\textit{FB}$ is the front-to-back power ratio. In this case, the modified expressions of \eqref{eq:J1GeneralFading} and \eqref{eq:J2GeneralFading} are
\begin{IEEEeqnarray}{rCl}
	J_1^\textit{BF}(x,y)&=& -\frac{y\tau  N_0\kappa^\alpha}{N_t}\left(\frac{x}{\lambda_{\text{B}}}\right)^{\frac{\alpha}{2}} \IEEEeqnarraynumspace \IEEEyesnumber\IEEEyessubnumber \label{eq:J1BeamForming}\\
	\displaystyle J_2^\textit{BF}(x,y)&=& -\pi x \left[1+\frac{1}{2\pi}\int\limits_{0}^{2\pi}I_\alpha\left(\frac{y A(\theta) P_I}{N_tW_I}\right)\, \mathrm{d}\theta\right] \IEEEeqnarraynumspace \IEEEyessubnumber \label{eq:J2BeamForming}
\end{IEEEeqnarray}
where the interference is evaluated over the continuous interval $\theta \in \left[0,2\pi\right]$.
\section{Numerical Results} \label{sec:results}
Numerical results illustrating the proposed theoretical framework are presented in this section. Regarding the adopted path loss model from \eqref{eq:PLmodel}, different values of the exponent $\alpha$ that reflect various propagation conditions are examined and the BS antenna height is assumed to be $L=30\text{m}$. The respective path loss constant is $\kappa=8.38\text{m}^{-1}$, defined in such a way that for $\alpha=3.5$ the obtained path losses are similar to the ones from the propagation prediction COST-Hata model for a medium sized city \cite{Cost231Hata}. Furthermore, a uniform traffic distribution with density $T_0=10~\text{Mbps}/\text{km}^2$ is considered, but other traffic profiles can easily be studied as well. The SNR gap is set equal to $\gamma_\textit{eff}=1$, the inverse cdf of fading is $F_h^{-1}\left(1-c\right)=1$, the noise density is $N_0=-174 \frac{\text{dBm}}{\text{Hz}}$ and unless otherwise stated, $N_K=10$ user groups are considered, where the different groups are formed by dividing the cell radius into equally spaced intervals.

\subsection{Single Cell}
\begin{figure}[!t]
\centering
\includegraphics[width=1\linewidth]{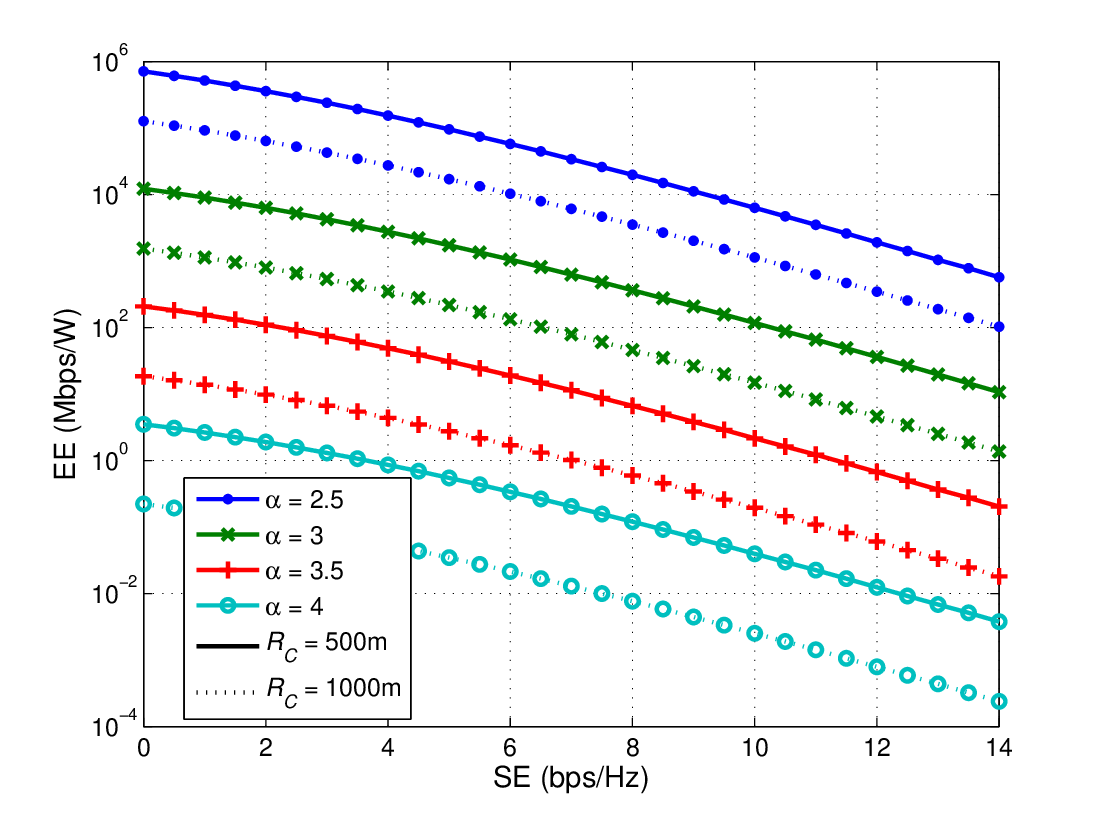}
\vspace{-10pt}
\caption{Single cell SE-EE trade-off from $(P_{\protect{\ref{pr:repartition}}})$ as a function of the path loss exponent and the cell radius.}
\label{fig:SingleCell}
\end{figure}

Firstly, \figurename{\ref{fig:SingleCell}} presents the SE-EE trade-off from $(P_{\ref{pr:repartition}})$ as a function of the path loss exponent $\alpha$ and the cell radius $R_C$. Similar behavior is observed for all the curves, where EE is a monotonically decreasing function of SE. This result is obviously expected, since SE receives higher values for less total bandwidth $W$, but at the same time, the needed transmit power becomes higher leading to inferior EE. Moreover, one can observe that EE is higher for better propagation conditions, i.e. lower values of the exponent $\alpha$, and it also increases for smaller cells. For example, looking at the case of $\alpha=3.5$ and $\textit{SE}=6 ~\text{bps}/\text{Hz}$, the EE is almost $19 ~\text{Mbps}/\text{W}$ for $R_C=500\text{m}$ and drops to $1.7~\text{Mbps}/\text{W}$ for $R_C=1000\text{m}$.

\begin{figure}[!t]
\centering
\includegraphics[width=1\linewidth]{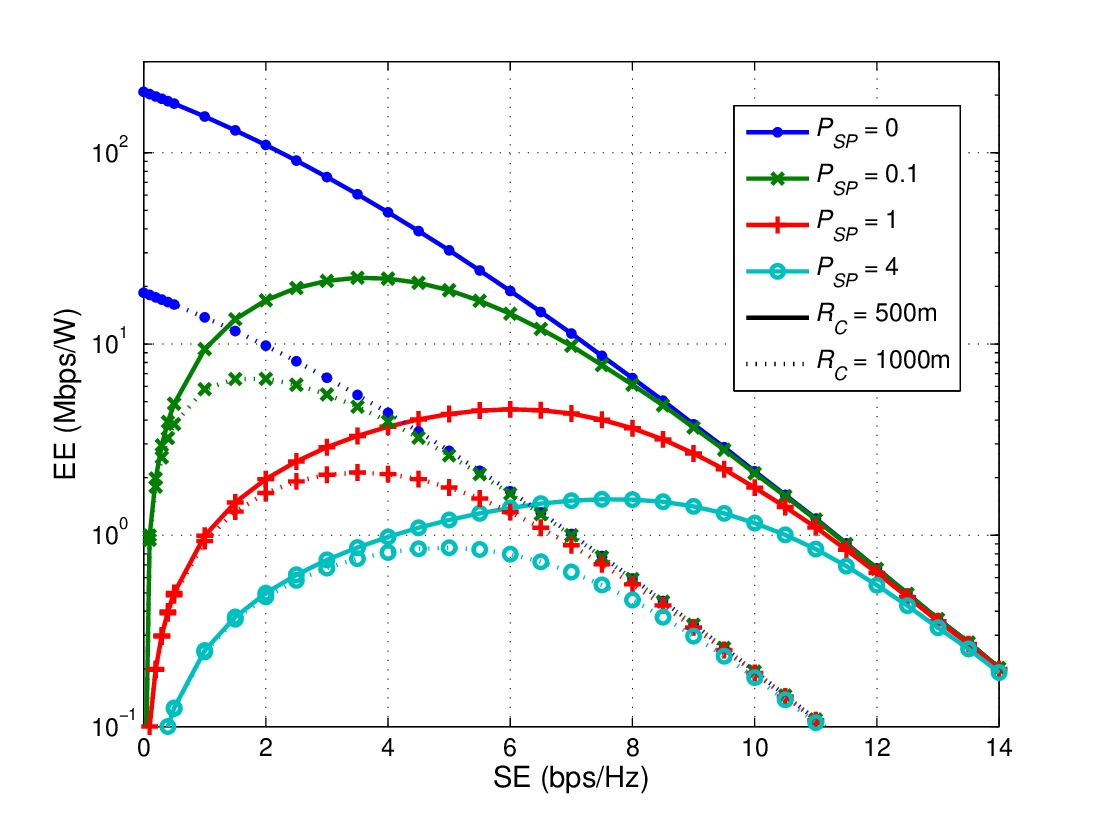}
\vspace{-10pt}
\caption{Single cell SE-EE trade-off from $(P_{\protect{\ref{pr:P_SP}}})$ as a function of the processing power factor and the cell radius, ($P_{SP}$ in W/MHz).}
\label{fig:SingleCell_PSP}\vspace{-10pt}
\end{figure}

The shape of the SE-EE trade-off is significantly changed and the relation is no longer monotonic when the signal processing power is taken into account in $(P_{\ref{pr:P_SP}})$, as \figurename{\ref{fig:SingleCell_PSP}} shows for $\alpha=3.5$. Specifically, since the processing power increases for higher bandwidth, it gradually becomes dominant in terms of consumed energy and after a critical point, EE starts to decrease by further bandwidth expansion. For instance, for processing power factor $P_{SP}=1~\text{W}/\text{MHz}$ and $R_C=500\text{m}$, EE is equal to $\lbrace1.8, 4.6, 2.0, 0.93\rbrace ~\text{Mbps}/\text{W}$ for decreasing SE with respective values $\lbrace10, 6, 2, 1\rbrace ~\text{bps}/\text{Hz}$. Eventually, the curves for different cell sizes become identical, indicating that at this extreme region the total power is mainly consumed by the signal processing and not the transmission. Note that for $P_{SP}>0$, the remaining part of the curves left from each critical point does not represent Pareto optimal solutions.

\begin{figure}[!t]
\centering
\includegraphics[width=1\linewidth]{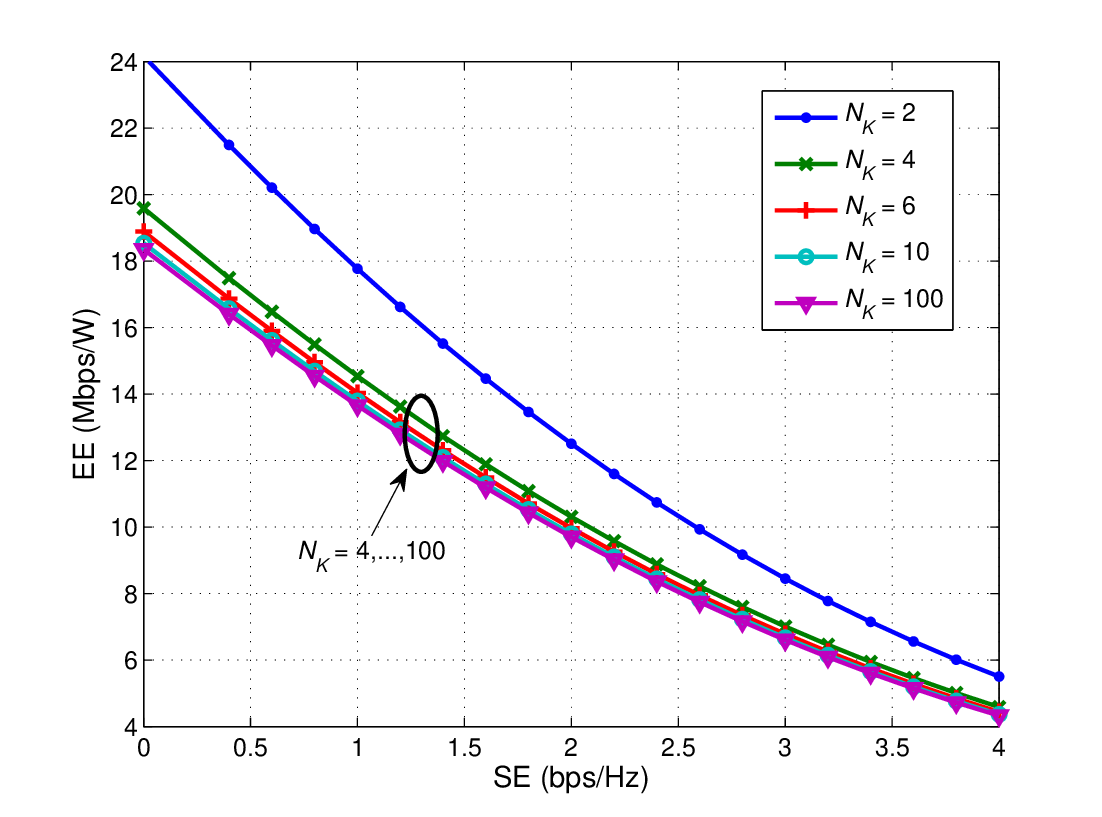}
\vspace{-10pt}
\caption{Convergence of SE-EE trade-off from $(P_{\protect{\ref{pr:repartition}}})$ with respect to the total number of user groups.}
\label{fig:Convergence}
\end{figure}

The next numerical results concern another important aspect of the traffic repartitioning approach. Due to the fact that the BS resources are allocated to users according to the central attenuation of their group and not their individual ones, there is an inevitable approximation error in this process. In this context, the convergence of the SE-EE curves from $(P_{\ref{pr:repartition}})$ with respect to the number of traffic partitions $N_K$ is illustrated in \figurename{\ref{fig:Convergence}}, for parameters $\alpha=3.5$ and $R_C=1000\text{m}$. One can see from this figure that the curves converge rapidly for the case of uniform traffic distribution and for the adopted value of ${N_K=10}$, the relative difference from  ${N_K=100}$ that approaches the discrete case is negligible. Nevertheless, if a scenario with specific user locations is studied instead of a continuous distribution, the optimal resource allocation from $(P_{\ref{pr:singleCell}})$ can be used to provide the exact SE-EE trade-off curves.

\begin{figure}[!t]
\centering
\includegraphics[width=1\linewidth]{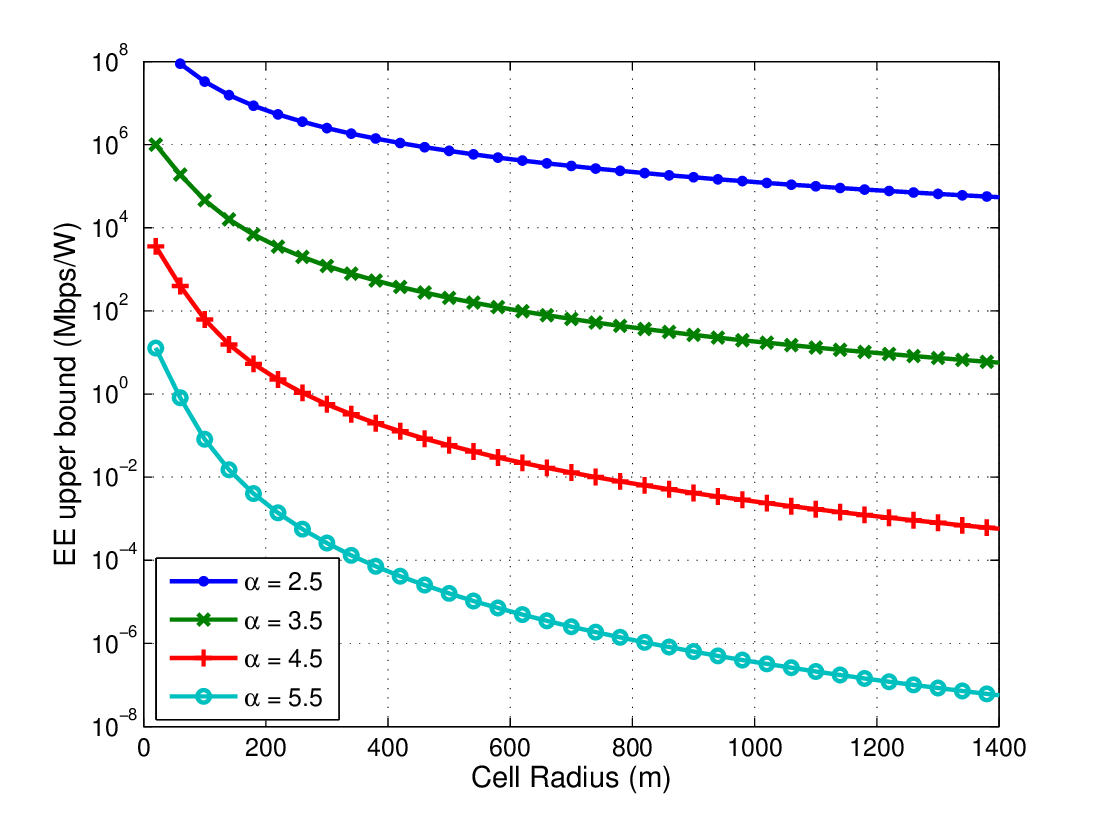}
\vspace{-10pt}
\caption{EE upper bound $\textit{EE}^*$ from \protect{\eqref{eq:EE_contLowSNR}} in the low SNR regime as a function of the path loss exponent and the cell radius.}
\label{fig:ContLowSNR}
\end{figure}

Concluding the results for the single cell scenario, the explicit EE upper bound $\textit{EE}^*$ from \eqref{eq:EE_contLowSNR}, achieved in the low SNR regime for $W\rightarrow \infty$, is presented in \figurename{\ref{fig:ContLowSNR}} as a function of the path loss exponent and the cell radius. The values in this figure actually represent the starting points of the curves in \figurename{\ref{fig:SingleCell}}, i.e. the highest EE as SE tends to zero. As expected, the EE upper bound is higher for better propagation conditions and smaller cells, where the needed transmit power is lower. It should be pointed out here that the range of values for $\textit{EE}^*$ is very large. In particular, for $R_C=500\text{m}$,  $\textit{EE}^*$ is equal to $206.1 ~\text{Mbps}/\text{W}$ for $\alpha=3.5$, dropping by a factor of approximately $3553$ times to $58 ~\text{Kbps}/\text{W}$ for $\alpha=4.5$. In such environments of high path loss exponents, smaller cells along with other techniques should be investigated as potential solutions for an energy efficient network.

\subsection{Multi-Cell Scenario}
We now consider the extension of the previous framework to a multi-cell scenario. The provided results are focused on the most tractable case with Rayleigh fading on both signal and interfering links, but can easily be extended to all the other presented mathematical models. Specifically, the resource allocation is performed according to the solution of $(P_{\ref{pr:multicell}})$, where the distribution of $Y_u$ is given from \eqref{eq:Ydistr} and the traffic groups are formed following \textit{Remark} \ref{rm:multicell} and \eqref{eq:T_k_Multicell} therein. Besides the previously adopted parameters, the fading mean value is equal to $1/\tau=1$ and the interfering BSs are assumed to have total available bandwidth $W_I =10~\text{MHz}$ and total transmit power $P_I =20~\text{W}$. The BS density, unless treated as a variable, is set equal to $\lambda_{\text{B}}=1 ~\text{point}/\text{km}^2$, while the users with $\textit{SINR}<-15\text{dB}$ are assumed to be in outage. 

\begin{figure}[!t]
\centering
\includegraphics[width=1\linewidth]{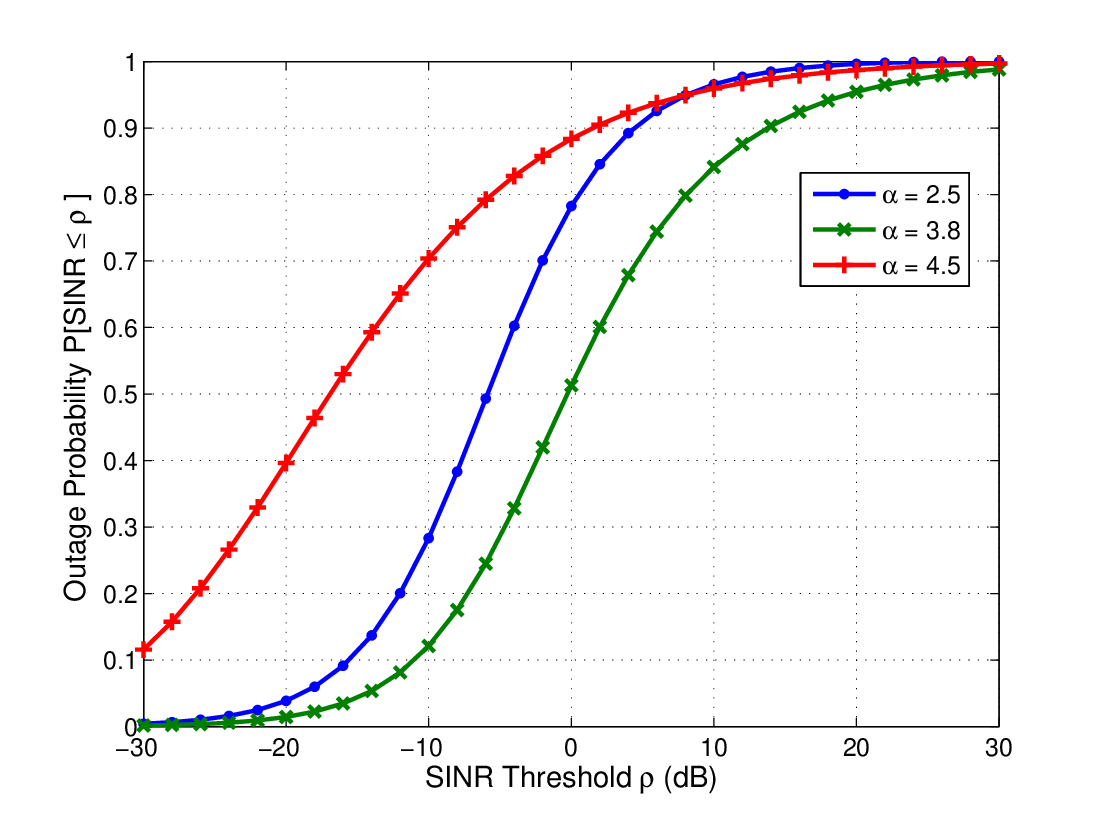}
\vspace{-10pt}
\caption{SINR cdf as a function of the path loss exponent for a user $u$ with $p_u/w_u=2\text{W}/\text{MHz}$.}
\label{fig:MultiCellcdfSINR}
\end{figure}
\begin{figure}[!t]
\centering
\includegraphics[width=1\linewidth]{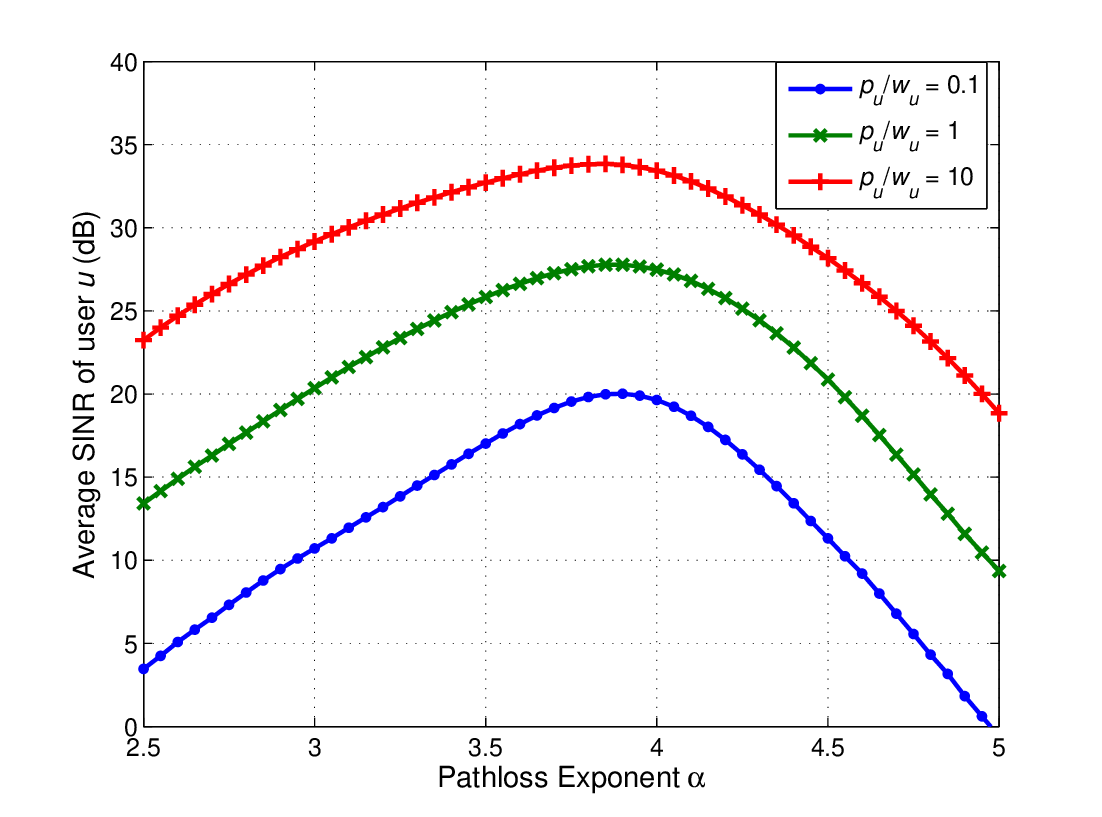}
\vspace{-10pt}
\caption{Average SINR as a function of the path loss exponent for a user $u$ with allocated resources $\left(w_u,p_u\right)$ (ratio $p_u/w_u$ in $\text{W}/\text{MHz}$).}
\label{fig:MultiCellAvgSINR}
\end{figure}

Before the obtained SE-EE curves are depicted, it is helpful to firstly examine the achieved SINR distribution. For this reason, \figurename{\ref{fig:MultiCellcdfSINR}} presents the SINR cdf for different values of the exponent $\alpha$ and \figurename{\ref{fig:MultiCellAvgSINR}} shows its average value, given by $\mathbb{E}\left[\textit{SINR}_u\right]=\int_{0}^{\infty}\mathbb{P}\left[\textit{SINR}_u>\rho \right]\, \mathrm{d} \rho$, as a function of $\alpha$ for different ratios of allocated resources $\left(w_u,p_u\right)$. It is interesting to notice from both figures that, as described in \cite{TsilimantosBCG}, the SINR is not a monotonic function of $\alpha$. Indeed, for all the curves in \figurename{\ref{fig:MultiCellAvgSINR}}, it initially increases until a certain point and then decreases for higher values of $\alpha$. This behavior is based on the nature of \eqref{eq:Ydistr} and specifically on the two functions $J_1(x,y)$ and $J_2(x,y)$ inside the integral. When $\alpha$ admits lower values, both the signal and the interference levels are increased, but eventually interference becomes dominant and $J_2(x,y)$ determines the result. On the other hand, if $\alpha$ is higher, even though the received interference is reduced, after a certain point the level of the desired signal starts to drop quickly and $J_1(x,y)$ turns out to be responsible for the degradation of SINR. Obviously, an optimal value of $\alpha$ lies between the two extremes, which in our case is around $\alpha=3.8$ for all the curves.

\begin{figure}[!t]
\centering
\includegraphics[width=1\linewidth]{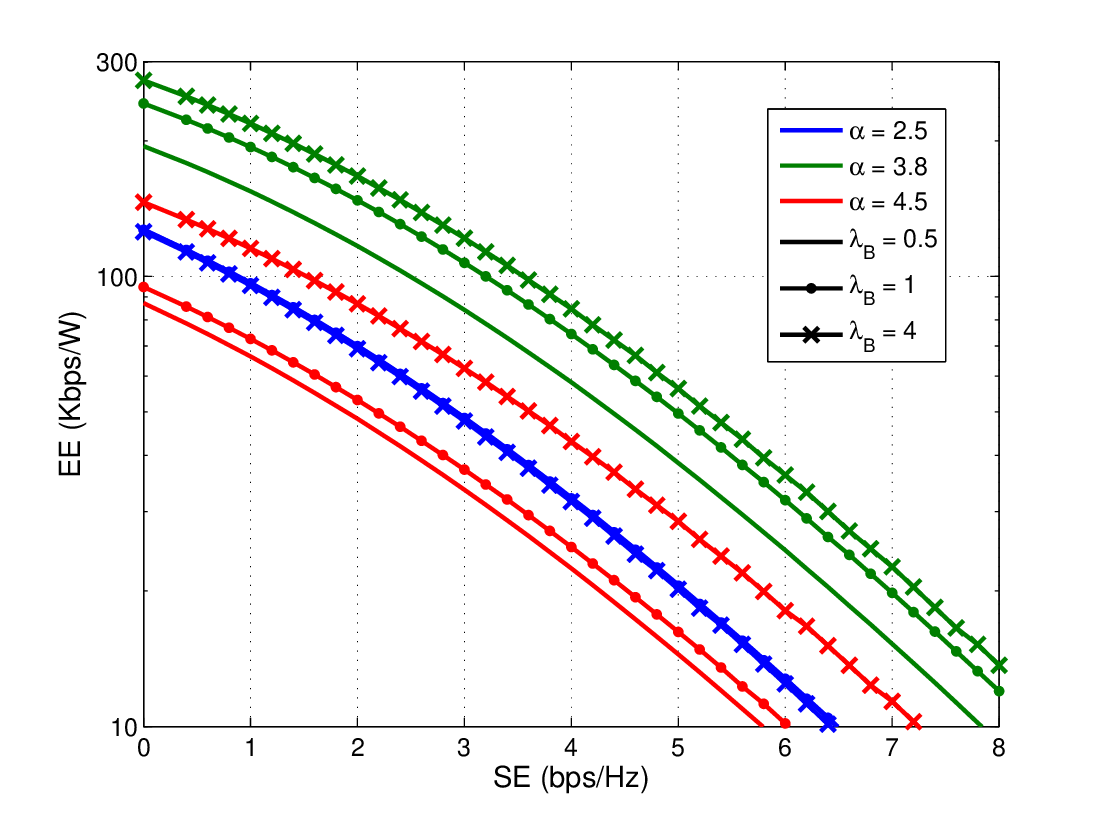}
\vspace{-10pt}
\caption{Multi-cell SE-EE trade-off from $(P_{\protect{\ref{pr:multicell}}})$ as a function of the path loss exponent and the BS density ($\lambda_\text{B}$ in $\text{points}/\text{km}^2$).}
\label{fig:MultiCell}
\end{figure}

The respective SE-EE trade-off from $(P_{\ref{pr:multicell}})$ as a function of the exponent $\alpha$ and BS density is shown in \figurename{\ref{fig:MultiCell}}. Similarly to the single cell case, EE is a monotonically decreasing function of SE where for the reasons already explained, the best performance is achieved for the intermediate value of $\alpha=3.8$. As expected, EE is better for smaller cells, i.e. higher values of $\lambda_{\text{B}}$, but this trend is not followed for $\alpha=2.5$ where EE remains almost the same. The reason behind this result is that the function $J_2(x,y)$, which is dominant for $\alpha=2.5$, does not depend on $\lambda_{\text{B}}$ according to \eqref{eq:J2Rayleigh}, due to the fact that the variation of the required signal power for different cell sizes is counter-balanced by the respective variation of the received interference. One should also observe that the estimated EE is quite low, indicating for example a value of $147.6~\text{Kbps}/\text{W}$ for $\textit{SE}=2 ~\text{bps}/\text{Hz}$, $\alpha=3.8$ and $\lambda_{\text{B}}=1 ~\text{point}/\text{km}^2$. Although this is a quite pessimistic value, since the currently adopted theoretical model does not incorporate the gain of directional antennas and the reduced interference due to the BS sectors, the significant difference from the respective results of the single cell scenario highlights the importance of interference management in multi-cell environments in terms of EE.

\begin{figure}[!t]
\centering
\includegraphics[width=1\linewidth]{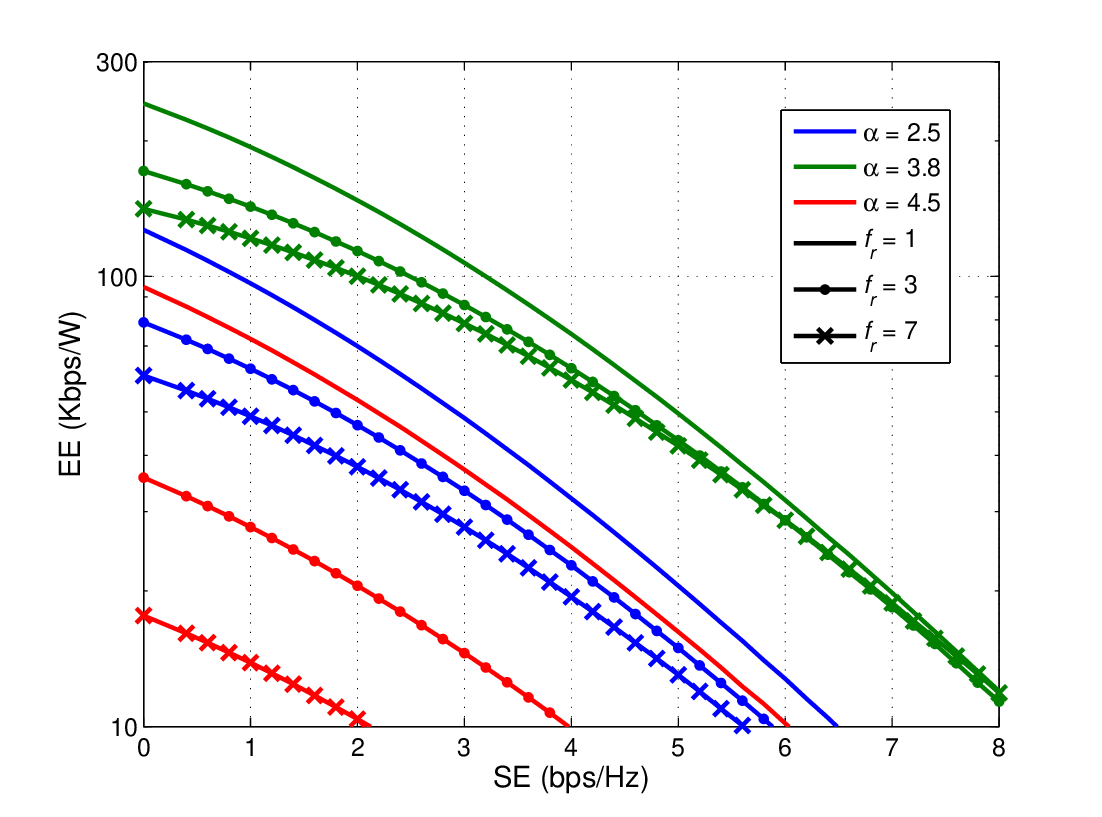}
\vspace{-10pt}
\caption{Multi-cell SE-EE trade-off from $(P_{\protect{\ref{pr:multicell}}})$ as a function of the path loss exponent and the frequency reuse factor.}
\label{fig:MultiCellFR} \vspace{-10pt}
\end{figure}
\begin{figure}[!t]
\centering
\includegraphics[width=1\linewidth]{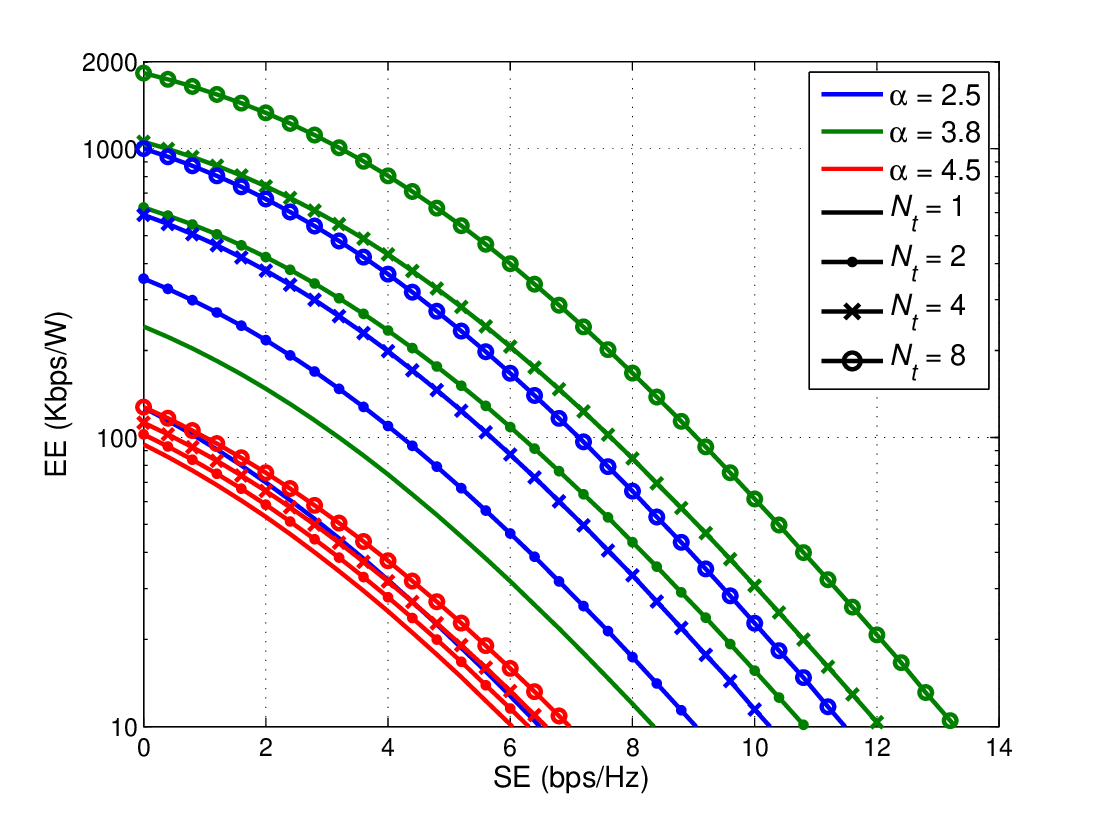}
\vspace{-10pt}
\caption{Multi-cell SE-EE trade-off from $(P_{\protect{\ref{pr:multicell}}})$ as a function of the path loss exponent and the number of BS transmit antennas.}
\label{fig:MultiCellBF}
\end{figure}

Along this line of thought, the impact of the interference mitigation techniques on the SE-EE trade-off is examined in the next two figures. Firstly, the utilization of the frequency reuse factor is studied and illustrated in \figurename{\ref{fig:MultiCellFR}}, where perhaps counter-intuitively, one can see that EE is a decreasing function of factor $f_r$. For instance, for $\textit{SE}=2 ~\text{bps}/\text{Hz}$ and $\alpha=3.8$, EE is equal to $\lbrace147.6, 113.8, 100\rbrace  ~\text{Kbps}/\text{W}$ for increasing $f_r$ with respective values $\lbrace1,3,7\rbrace$. The reason for this result is that although the SINR is improved for higher values of $f_r$ due to the reduced received interference, the BS bandwidth resources are less, since $\sum_{k \in \mathcal{K}}W_K=W/f_r$ and the needed transmit power must be increased, leading to an overall deterioration of EE. This conclusion comes into agreement with a similar finding in \cite{BaccelliCellular}, where the optimal user mean rate is achieved for $f_r=1$.

Finally, the impact of beamforming on the SE-EE trade-off is presented in \figurename{\ref{fig:MultiCellBF}} for different values of the exponent $\alpha$ and the number $N_t$ of antennas, while $G_\textit{FB}=20\text{dB}$. In this case, the received interference is reduced and at the same time EE is significantly increased for all curves, especially in the region where $\alpha$ is not too high and the function $J_2(x,y)$ is dominant. Using the same numerical example of $\textit{SE}=2 ~\text{bps}/\text{Hz}$ and ${\alpha=3.8}$, EE is equal to $\lbrace147.6, 421.1, 740.5, 1329\rbrace  ~\text{Kbps}/\text{W}$ for a number of antennas $N_t=\lbrace1,2,4,8\rbrace$ respectively. This verifies the critical role that advanced antenna techniques can play in order to achieve network performance enhancements.

\section{Conclusions}
This paper has introduced a simple theoretical framework for studying the achieved SE-EE trade-off in cellular networks, leading to tractable analytical results. An optimal resource allocation formalism has been presented for both single and multi-cell environments and the complexity of the underlying optimization problem has been significantly reduced by applying a proposed traffic repartition scheme that limits the task of resource allocation to group of users with similar channel conditions. Moreover, a theoretical EE upper bound has been provided for the special case of low SNR and an extension of the theoretical model has been examined by taking into account the signal processing power.

Our results have shown that in the case of a single cell, as expected, EE is higher for smaller cells and lower path loss exponent and that EE is no longer a monotonically decreasing function of SE when the signal processing power is included in the evaluation of the total power consumption. The traffic repartition approach also proves to be very accurate with a small number of user groups. EE is substantially lower in the multi-cell scenario, where an optimal value of the path loss exponent exists. The small cells are still preferred in terms of EE, especially for environments with higher path loss exponents. Regarding the interference mitigation methods, beamforming proves to be a useful technique to improve EE, but on the other hand, frequency reuse fails to achieve the same goal due to the reduction of the available bandwidth.

This work encourages further research towards more sophisticated power, channel and traffic models that will lead to more accurate numerical results. Also of interest are more advanced multi-cell scenarios and emerging deployment concepts such as heterogeneous networks, a promising approach to cope with the challenging issue of developing green cellular networks.
\appendices

\section{Proof of Lemma \ref{lem:solution}}
\label{app:lemma}

Summing \eqref{eq:w_u} over $u \in \mathcal{U}$ allows us to use the second condition of \eqref{eq:LamdaDeriv} and form the following function:
\begin{equation}
f(\lambda)=W- \displaystyle \sum\limits_{u \in \mathcal{U}}\frac{T_u \ln 2}{1+W_0\left(\frac{1}{e}\left[\frac{\lambda\ell_u F_h^{-1}(1-c)}{\gamma _\textit{eff}N_0}-1\right]\right)}\, .
\label{eq:function_l}
\end{equation}

Firstly, one should notice that the desired value of $\lambda$ is the solution of the equation $f(\lambda)=0$. Moreover, since the real-valued Lambert function $W_0$ is monotonic, it is easily seen that the function $f$ is also monotonic. Finally, a way to bound $\lambda$ from both sides is to assume that all users experience the same attenuation equal to the minimum $\ell_{m}$ or the maximum $\ell_{M}$, with $\ell_{m}>\ell_{M}$, and then solve $f(\lambda)=0$. As a result, $\lambda$ should lie between these two values, and specifically
\begin{equation}
\begin{array}{c}
	\displaystyle \lambda \in \left[\frac{q}{\ell _m},\frac{q}{\ell _M}\right], ~ \text{with} \\[10pt] \displaystyle q\triangleq \frac{\gamma _\textit{eff}N_0 \left(ze^{z+1}+1\right)}{F_h^{-1}(1-c)},~ z\triangleq  \frac{\ln 2}{W}\sum\limits_{u \in \mathcal{U}}T_u -1	\, .
	\end{array}
	\label{eq:lambda}
\end{equation}

\section{Proof of Theorem \ref{th:repartition}}
\label{app:th2}

According to \textit{Theorem} \ref{th:single}, for any user $u\in \mathcal{U}_k$
\begin{IEEEeqnarray}{rCl}
w_u^*=c_1T_u&,&~c_1 \triangleq \frac{\ln 2}{1+W_0\left(\frac{1}{e}\left[\frac{\lambda\ell_k F_h^{-1}(1-c)}{\gamma _\textit{eff}N_0}-1\right]\right)} \IEEEyesnumber \IEEEyessubnumber \IEEEeqnarraynumspace
\label{eq:proof1 w_u}\\
p_u^*=c_2T_u&,&~c_2 \triangleq \frac{\gamma _\textit{eff} N_0}{\ell_k F_h^{-1}(1-c)}\left(2^\frac{1}{c_1}-1\right)c_1 \IEEEyessubnumber \IEEEeqnarraynumspace
\label{eq:proof1 p_u}
\end{IEEEeqnarray}
where $c_1$ and $c_2$ are the constants defined here and $\ell_k$ is the common attenuation of the set $\mathcal{U}_k$ from \eqref{eq:groups}. Summing both sides of \eqref{eq:proof1 w_u}-\eqref{eq:proof1 p_u} over $u \in \mathcal{U}_k$ leads to
\begin{IEEEeqnarray}{rCl}
W_k=c_1T_k & \xrightarrow{\eqref{eq:proof1 w_u}}& w_u^*=\frac{T_u}{T_k} W_k  \IEEEyesnumber\IEEEyessubnumber \label{eq:proof2 w_u}\\
P_k=c_2T_k & \xrightarrow{\eqref{eq:proof1 p_u}}& p_u^*=\frac{T_u}{T_k} P_k  \IEEEyessubnumber
\label{eq:proof2 p_u}
\end{IEEEeqnarray}
and this completes the proof.

\section{Proof of Lemma \ref{lem:lowSNR}}
\label{app:lowSNR}
A common approach to simplify the problem is to linearize it by approximating the natural logarithm with the help of the Taylor series, which yields
\begin{equation}
\ln \left(1+x\right)=x-\frac{x^2}{2}+\frac{x^3}{3}-\dots \simeq x ~~\forall \left|x\right| \ll 1 \, .
\label{eq:Taylor}
\end{equation}

By writing the first constraint of $(P_{\ref{pr:singleCell}})$ as a function of $T_u$ and applying this property, we obtain
\begin{equation}
T_u=w_u\log_2 \left(1+\frac{p_u\ell_uF_h^{-1}(1-c)}{\gamma _\textit{eff}N_0w_u}\right)\simeq\frac{p_u\ell_uF_h^{-1}(1-c)}{\gamma _\textit{eff}N_0 \ln 2}
\label{eq:T_u lowSNR}
\end{equation}
which is linear in the received power and as expected insensitive to the bandwidth.
The required total transmit power is then found by solving \eqref{eq:T_u lowSNR} with respect to $p_u$ and summing over $u \in \mathcal{U}$. Finally, the expression for EE from \eqref{eq:EE} gives the upper bound \eqref{eq:EE_tot lowSNR}, as stated in Lemma \ref{lem:lowSNR}.
\IEEEtriggeratref{36}
\bibliographystyle{IEEEtran}
\bibliography{IEEEabrv,reflist}

\end{document}